\documentclass{aa}  
\usepackage[utf8]{inputenc}
\usepackage{rotating}
\usepackage{float}
\usepackage{longtable}
\usepackage{lscape}
\usepackage{array}
\usepackage{graphicx}
\usepackage{booktabs}
\usepackage{tabularx}
\usepackage{subcaption}
\usepackage{stfloats}
\usepackage{adjustbox}
\usepackage{txfonts}
\usepackage[colorlinks=true, linkcolor=blue, citecolor=blue, urlcolor=blue]{hyperref}
\hypersetup{pdfborder={0 0 0}}
\makeatletter
\renewcommand{\fnum@figure}{\textbf{\textcolor{blue}{Fig.~\thefigure}}}
\makeatother
\makeatletter
\renewcommand{\fnum@table}{\textbf{Table~\thetable}}
\makeatother

\begin{document}

   \title{Re-visiting the Canis Major star-forming region with Gaia data release 3 data}

   \author{S. Nogueira-Silva
          \inst{1}
          \and
          P.A.B. Galli
          \inst{2}
          \and
          J. Olivares
          \inst{3}
          \and
          H. Bouy 
          \inst{4,5}
            \and
          B. P. Popowicz
          \inst{2}
          \and
          P. M. Merino
          \inst{2}
                \and
          T. Santos-Silva \inst{2}
                \and
          J.~Gregorio-Hetem \inst{2}
                \and
           N. Miret-Roig\inst{6,7,8}
                \and
          J. Alves \inst{6}
          }

    \institute{Núcleo de Astrofísica, Universidade Cidade de São Paulo, R. Galvão Bueno 868, Liberdade, 01506-000 São Paulo, SP, Brazil\\
                \email{sabrinavitoria53@gmail.com}
        \and
            Instituto de Astronomia, Geofísica e Ciências Atmosféricas, Universidade de São Paulo, Rua do Matão, 1226, Cidade Universitária,
05508-090 São Paulo-SP, Brazil\\
        \and
             Departamento de Inteligencia Artificial, Universidad Nacional de Educación a Distancia (UNED), c/Juan del Rosal 16, 28040 Madrid, Spain\\
        \and
            Université de Bordeaux, Lab. d’Astrophysique de Bordeaux,CNRS, B18N, Allée Geoffroy Saint - Hillaire, 33615 Pessac,France.\\
        \and
            Université de Bordeaux, Institut universitaire de France (IUF), 1 rue Descartes, 75231 Paris CEDEX 05\\
        \and
            University of Vienna, Department of Astrophysics, Türkenschanzstraße 17, 1180 Vienna, Austria. 
        \and 
            Dep.~de Física Quàntica i Astrofísica (FQA), Univ.~de Barcelona (UB), Martí i Franquès, 1, 08028 Barcelona, Spain
        \and
            Institut de Ciències del Cosmos (ICCUB), Univ.~de Barcelona (UB), Martí i Franquès, 1, 08028 Barcelona, Spain \\
            }

\date{Received ***; accepted **}

\abstract
{The Canis Major (CMa) star-forming region, a remote molecular cloud complex within the recently discovered Radcliffe Wave, remains under-explored in the literature.}
{We revisit the stellar census in the CMa region, characterizing its stellar population, kinematics, and age using recent astrometric and photometric data from the third data release of the Gaia space mission (Gaia DR3).}
{We conducted a membership analysis of Gaia DR3 sources across a 16 deg$^2$ field encompassing the youngest subgroups in CMa. This new stellar census, combined with spectroscopic observations, allowed us to investigate the structure, kinematics, and age of this region.}
{We identified 1 531 objects as members of the CMa region, confirming 401 previously known members and introducing 1\,130 new candidate members. These objects have magnitudes ranging from 10 to 18 mag in the G band from Gaia DR3. We identified two subgroups of CMa stars in our sample labelled as Cluster A and Cluster B. They are located at roughly the same distance ($d_{A} = 1\,150^{+79}_{-88}$~pc and $d_{B} = 1\,183^{+103}_{-108}$~pc) and exhibit similar space motions that can be derived thanks to the precise radial velocities obtained in this study. The subgroups have a mean isochronal age of about 2-3~Myr. However,  based on infrared photometry we show that Cluster A has a higher fraction of disc-bearing stars suggesting that it could be somewhat younger than Cluster B. }
{Our analysis provides new insights into the stellar population of the Canis Major region, by identifying new members, characterizing their kinematics, and assessing their evolutionary stages. Future studies incorporating additional data from upcoming Gaia data releases, multi-wavelength and high-resolution spectroscopic observations will be essential to further advance our understanding of the history of star formation in this region.}

\keywords{radial velocities -- kinematics and dynamics  -- astrometry -- parallaxes -- star formation}

   \maketitle
%
\section{Introduction}

The Canis Major (CMa) star-forming region is a remarkable laboratory for studying the formation and evolution of stars. Located approximately 1.2 kpc from the Sun \citep{2008hsf2.book....1G}, this region is home to more than 200 luminous B-type stars and a few late-type O stars. First identified as a stellar association by \citet{ambartsumian1947evolution}, subsequent studies \citep{1978ApJ...223..471H} have revealed that CMa hosts a predominantly young stellar population (more than a hundred young stars), rich in pre-main-sequence stars with an average age of just 0.5 Myr. This cosmic nursery is closely linked to the reflection nebula CMa R1, which is complemented by three prominent HII regions—Sh 2-292, Sh 2-296, and Sh 2-297 \citep{1978ApJ...223..471H}—and contains over a dozen known open clusters \citep{reffert}, establishing it as a dynamic hub of star formation.

The region also includes six dense dark clouds LDN 1653–LDN 1658 \citep{2005PASJ...57S...1D}, which obscure parts of CMa and contain numerous embedded stars, indicating ongoing star formation. Initially referred to as I Canis Majoris by \citet{ambartsumian1947evolution}, this area was later reclassified as CMa OB1 by \citet{ruprecht1966classification}. The relationship between CMa OB1 and CMa R1 has since been clarified, with CMa R1 identified as a reflection nebula embedded within the broader CMa OB1 association. Notably, \citet{1974b} confirmed that the majority of stars studied by \citet{1968AJ.....73..233R} in CMa R1 are indeed members of CMa~OB1.

X-ray studies have also contributed to the understanding of CMa. The first wide-field X-ray study of the young stellar population associated with CMa R1 was performed by ROSAT data, revealing the previously unknown older, fainter low-mass stellar population \citep{2009A&A...506..711G}. Using XMM-Newton, \citet{2018} analysed the Sh 2-296 nebula, identifying 58 members of the region, including 41 T Tauri stars and 15 additional pre-main-sequence objects. These studies revealed that half of the young stars in the region have masses below 1~$M_\odot$ and ages between 1 and 2 Myr. Despite these complexities, CMa remains an invaluable site for investigating young stellar populations and their interactions with the interstellar medium. 

In \citet{Fischer_2016}, young stellar objects (YSOs) in the CMa region were classified as Class I and Class II within a 100 deg² field. A total of 335 sources were identified as Class II, while 144 were classified as Class I. Class I objects correspond to an earlier evolutionary phase, in which the protostar remains deeply embedded within its natal envelope. In contrast, Class~II objects are more evolved, characterized by the presence of a prominent protoplanetary disc and reduced circumstellar obscuration. Based on these classifications, \citet{Fischer_2016} concluded that the stellar population in this region is predominantly Class II objects. This trend was also noticed by \citet{2015MNRAS.448..119F}, in the characterization of 41 T Tauri stars associated with the Sh 2-296 nebula. They found that half of the sample has ages  < 1–2 Myr, but only a small fraction (25\%) shows evidence of IR excess due to the presence of circumstellar discs.

The distance to the CMa region remains debated, with estimates varying from 1\,000 to 1\,300 pc \citep{10.1046/j.1365-8711.1999.02937.x,2019A&A...630A..90P, zucker}, depending on the objects studied. For instance, \citet{1974b} used 36 confirmed association members to derive an average distance of 1\,150 pc. Also, \citet{10.1046/j.1365-8711.1999.02937.x} analysed 165 stars brighter than magnitude 13 in the CMa R1 region, identifying 88 early-type candidate members with a colour excess \( E(B - V) = 0.16 \) mag, corresponding to a distance of approximately 1 kpc. 

In \cite{2019A&A...630A..90P}, the study focused on H$\alpha$ emitters detected by WISE, which are predominantly concentrated in star-forming regions. The authors identified 398 objects classified as H$\alpha$ emitters. The distance distribution for these objects spans the range of 1\,050 to 1\,350 pc, with a pronounced peak at 1\,185 pc. Additionally, using OB stars as reference points, they determined a median distance of 1\,282 pc. 

In \cite{zucker}, the authors focused on calculating precise distances to local molecular clouds using Bayesian inference. Their study included four molecular clouds in the CMa OB1 region, with distances ranging from 1\,169 to 1\,268 pc. 

\citet{Santos2021} identified four subgroups within the CMa region, designated as CMa05, CMa06, CMa07, and CMa08,  which consist of young stellar populations with ages between 10 and 20 Myr, located at distances ranging from 1\,000 to 1\,200 pc. Among these, CMa06 stands out due to its distinct proper motion distribution compared to the overlapping proper motions of the other subgroups. More recently, \citet{2024AJ....168..225D} analysed the distributions of molecular gas, based on the positions and velocities of independent structures revealed by the 12CO data. These gas structures  were combined with the differences in distances and motions of the YSOs to suggest a division of the CMa region into seven subregions. The distance found for the subregions ranges from 1\,080 pc to 1\,159 pc.

The CMa region is part of the Radcliffe Wave \citep{alves2020galactic}, a large-scale galactic structure composed of interconnected star-forming regions, extending over more than 2.7 kpc. Canis Major is located near the outer edge of this structure, making it one of the most distant regions along the wave, similar to the Cygnus-X complex \citep{alves2020galactic}.

Despite the wealth of available data, precise radial velocity measurements for this region remain scarce, as Gaia’s radial velocity data lack sufficient precision due to large uncertainties. Accurate kinematic measurements are therefore crucial for characterizing the region and understanding its relationship with the larger galactic structures.

With the recent third data release of the Gaia space mission \citep{2023A&A...674A...1G}, the CMa region can be reanalysed using high-precision astrometric and photometric data. Membership analysis methods allow us to characterize the stellar population, while the inclusion of unprecedented radial velocity data establishes CMa as a fresh starting point for this study, through the unique combination of Gaia DR3 data with ground-based spectroscopy. The primary goal of this work is to characterize the CMa region in terms of its stellar population, spatial structure, kinematics, and age.

This paper is structured as follows. In Sect.~\ref{section2}, we perform a new membership analysis of the CMa region centred around the youngest subgroups identified by \cite{Santos2021} based on Gaia DR3 data. In Section~\ref{subsection:radial} we describe our observations and procedure to derive precise radial velocity measurements of CMa stars from ground-based observations. In Sect.~\ref{section4}, we reassess key properties of the cluster, such as distance, kinematics, age, and spatial distribution, using our new sample of cluster members. Finally, we summarize our findings in Sect.~\ref{section5}.

\section{Membership analysis}\label{section2}

In this section, we outline our strategy to identify new members of the CMa region and confirm previous candidates. The methodology employed to select the most likely members of CMa is based on the methods developed by \cite{Sarro_2014} and \cite{Olivares_2019}. 

Our membership analysis is based on data from the Gaia DR3 catalogue. We downloaded the Gaia DR3 catalogue in the CMa region defined by $104^\circ < \alpha < 108^\circ \quad \text{and} \quad \nobreak{-13.4^\circ} < \delta < \nobreak{-10^\circ}$ that encompasses the youngest clusters identified by \cite{Santos2021}. The initial catalogue comprised 736\,632 sources. After downloading the data, we applied a cleaning process by imposing a 3$\sigma$ cut on the proper motion modulus, resulting in a final selection of 162\,331 sources in the field. The representation space (i.e. space of parameters) that we use in the membership analysis includes both astrometric and photometric features from the Gaia DR3 catalogue ($\mu_{\alpha}\cos\delta$, $\mu_{\delta}$, $\pi$, $G$, and $G-RP$). 

We identified probable CMa members by modelling the field and cluster populations. The field population was characterized with a Gaussian mixture model (GMM) applied to both astrometric and photometric spaces. The optimal number of Gaussian components was determined using the Bayesian information criterion (BIC). We tested models with the number of components ranging from 40 to 180. The model with 80 components yielded the lowest BIC value and was subsequently adopted to construct the field model. This field model was computed at the beginning and remained static throughout the analysis.

The cluster model is generated independently, combining a GMM in the astrometric space ($\mu_{\alpha}\cos\delta$, $\mu_{\delta}$, and $\pi$). The photometric model's mean is represented by a principal curve, following the cluster's sequence. Our method calculates membership probabilities for each source and classifies the sources as members or non-members based on a user-defined probability threshold, $p_{in}$. In this study, we tested threshold values ranging from 0.5 to 0.9 (corresponding to 50\% to 90\%).

Our membership analysis takes an initial list of candidate members to define the locus of the cluster in the space of parameters in the first iteration of the code. We used as the initial list of members the sample of cluster members for the youngest populations of CMa subgroups  (CMa05, CMa06, CMa07, and CMa08) provided by \cite{Santos2021}. This sample of candidate members was obtained from Gaia DR2 data. We updated this dataset with Gaia DR3 data and included it in our analysis.

From this point, the algorithm iteratively refined the list of members based on their membership probabilities. The algorithm iterated until convergence, which occurs when the membership list stabilises after successive iterations. Then, we generated synthetic data using the field and cluster models that were computed in our analysis as explained before. We determined the optimal probability threshold of our analysis after computing the true positive rate (TPR) and contamination rate (CR) of the sample as explained in \cite{Olivares_2019}.

\begin{table}[!h]
\centering
\caption{Results of our membership analysis with various probability thresholds.}
\label{table1}
\begin{tabular}{c c c c c} 
 \hline\hline
 $p_{in}$ & $p_{opt}$ & Members & TPR & CR\\ 
 \hline\hline
 0.6 & 0.939 & 1\,559 & 0.772  $\pm$  0.010 & 0.314  $\pm$  0.020 \\
 0.7 & 0.909 & 1\,860 & 0.821  $\pm$  0.003 & 0.304  $\pm$  0.001 \\ 
 0.8 & 0.858 & 1\,531 & 0.821  $\pm$  0.005 & 0.336  $\pm$  0.001 \\
 0.9 & 0.676 & 683 & 0.721 $\pm$  0.018 & 0.281  $\pm$  0.021 \\
 \hline
\end{tabular}
\vspace{0.5em}
\tablefoot{Parameter $p_{in}$ denotes initial membership probability, while $p_{opt}$ represents optimized probability threshold. Members indicates the final number of selected objects. TPR and CR correspond to the true positive rate and contamination rate, respectively.}
\end{table}

Table \ref{table1} summarizes the results for different user-defined probability thresholds, $p_{in}$. The results with $p_{in}$ = 0.5 were omitted from Table \ref{table1} due to convergence issues of the field model that are most likely caused by the high contamination in this solution. We note from Table \ref{table1} that the number of members retrieved in each solution decreases with the increasing $p_{in}$ threshold (the only exception is the solution obtained with $p_{in}$ = 0.7). It is also clear from Table \ref{table1} that a less restrictive $p_{in}$ value is often associated with a more conservative $p_{opt}$ (see, for example, the solutions obtained with $p_{in}$ = 0.6 and $p_{in}$ = 0.9). The closest match between these two probability thresholds occurs with $p_{in}$ = 0.8 (and $p_{opt}$ = 0.858).

As shown in Table \ref{table1}, the most extreme solutions ($p_{in}$ = 0.6 and $p_{in}$ = 0.9) exhibit the lowest TPRs and can therefore be excluded. We note that the faintest sources are missing in the solution with $p_{in}$ = 0.9, which prevents the detection of low-mass cluster members and the study of the initial mass function (IMF) in the faint end. When comparing the other two solutions, we note a more significant dispersion in the proper motion and parallax diagrams of the sources in the $p_{in}$ = 0.7 solution compared to the results obtained with $p_{in}$ = 0.8.

\begin{figure}[!h]
   \centering
   \includegraphics[width=0.49\textwidth]{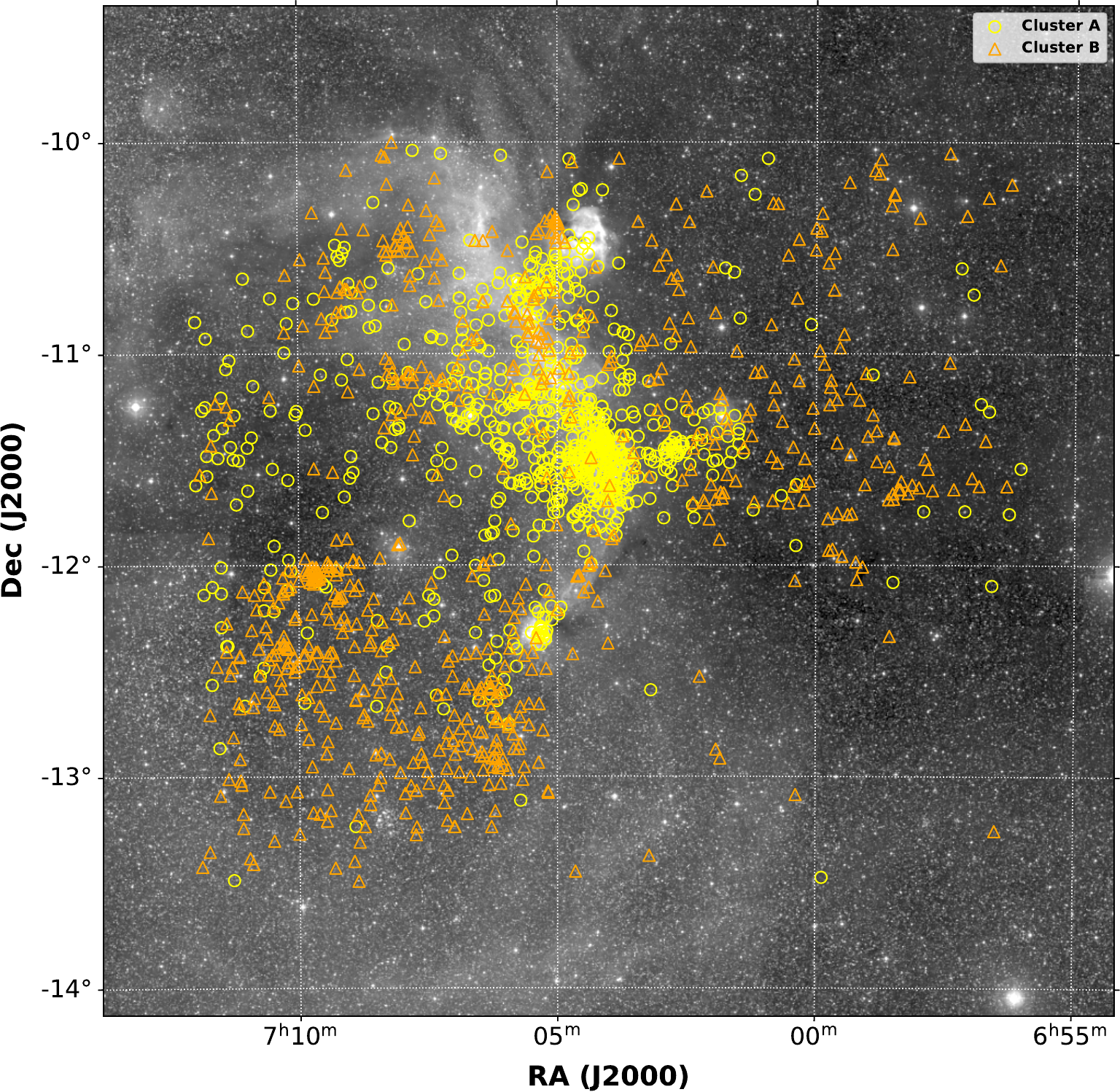}
   \caption{Spatial distribution of 1\,531 stars identified as members of CMa region, overlaid on Digitized Sky Survey 2 (DSS2) red-band image. The stars are colour-coded according to their subgroup membership: yellow symbols indicate stars belonging to Cluster A, while orange symbols represent stars assigned to Cluster B, as described in Section~\ref{subsec:subgroups}.}
   \label{ravsdec}
\end{figure}

We have therefore chosen to work with the solution obtained from $p_{in}$ = 0.8, as it provides the highest TPR and the smallest difference between the user-defined and optimal probability thresholds. Consequently, the final list of cluster members derived from our study includes 1\,531 stars (see Fig. \ref{ravsdec}). Table \ref{A1} lists all the 1\,531 cluster members identified in our membership analysis. In Table \ref{A2} we provide the membership probabilities of individual sources in the field for the solutions investigated in this study with different $p_{in}$ thresholds. 

\begin{figure*}[ht]
    \centering
    \includegraphics[width=0.85\textwidth]{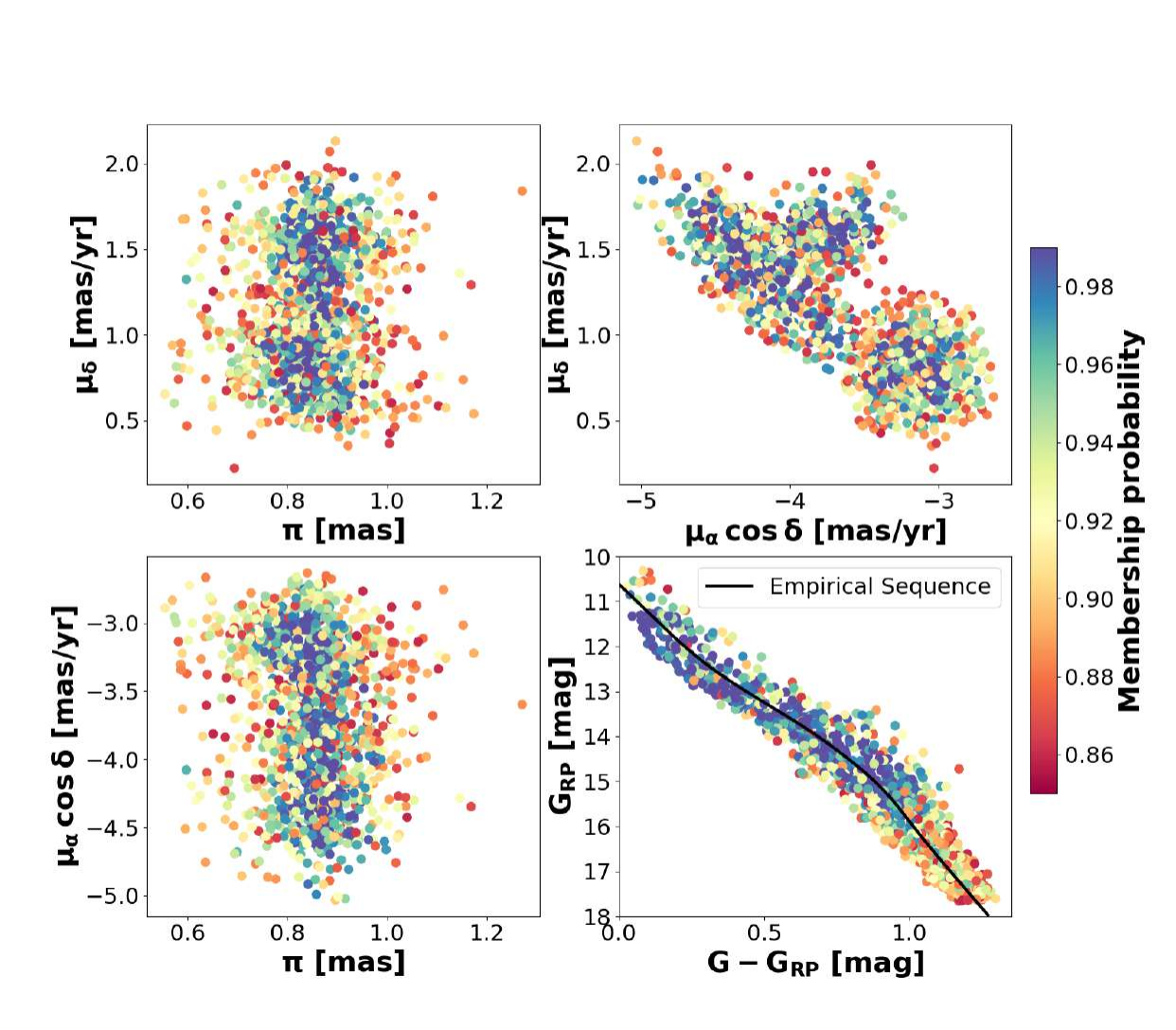}  
    \caption{Parallaxes and proper motions of 1\,531 stars identified in our membership analysis and colour-magnitude diagram of CMa sample found by our membership investigation. The empirical isochrone, derived in our analysis, is indicated by the solid black line. The individual membership probabilities of the stars are scaled from 0 to 1 and shown with different colours.}
    \label{all}
\end{figure*}

Figures \ref{ravsdec} and \ref{all} show the distribution of position, proper motion, and parallax of the sample of CMa stars selected in our analysis. It is apparent from Fig. \ref{all} that our sample of CMa stars consists of different subgroups (see Sect. \ref{subsec:subgroups} for more details). Moreover, we note that the more dispersed sources in the space of proper motions and parallax have the lowest membership probabilities.

\begin{figure}[ht]
   \centering
   \includegraphics[scale=0.35]{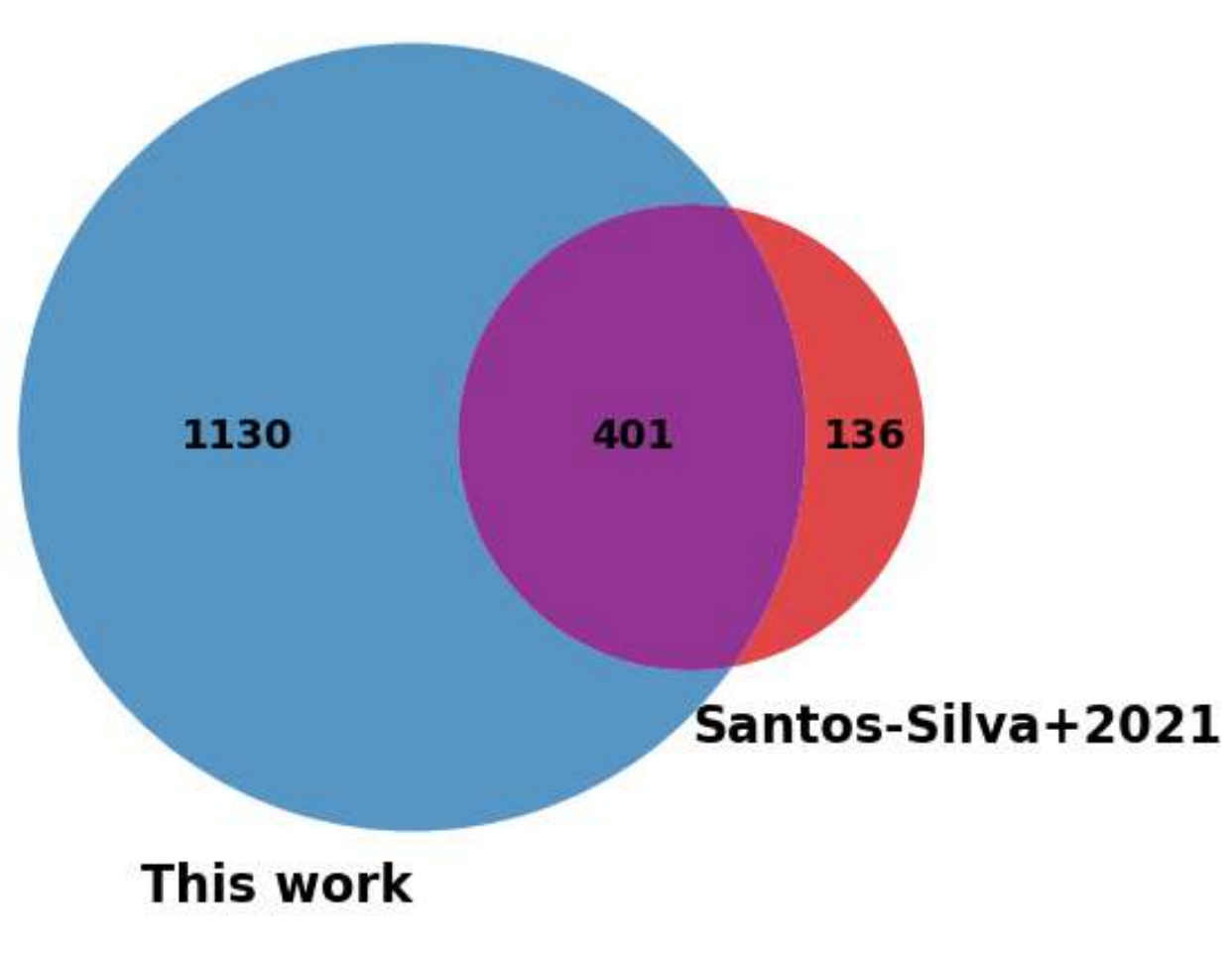}
   \caption{Venn diagram comparing the number of stars in common between our analysis and the previous study by \citet{Santos2021}.}
   \label{venn}
\end{figure}

Figure~\ref{all} also shows the colour–magnitude diagram (CMD) for the sources in our sample. The CMD displays apparent magnitudes from Gaia without any extinction correction applied. For extinction-corrected magnitudes, refer to Fig.~\ref{fig9}. As depicted, CMa stars primarily populate the magnitude range of 10 to 18~mag in the Gaia G band.  We observe a distinct magnitude cutoff, which becomes more pronounced as the membership probability threshold increases. For instance, at a threshold of $p_{in} = 0.8$, no sources fainter than 18 mag are found, indicating that objects at higher (fainter) magnitudes generally exhibit lower membership probabilities. Similarly, sources brighter than G = 10 mag are absent from our sample due to selection constraints. The brightest star included has a magnitude of G = 10.4~mag and corresponds to Gaia DR3 3046209991397371392, classified as a B5-type star.

Figure \ref{venn} presents a comparison of cluster members identified in this study with those of \cite{Santos2021}. Our membership analysis confirms 401 stars from the initial list of candidate members based on the sample identified by \cite{Santos2021}, while rejecting 136 sources from that list. We retrieved 296 objects from the CMa06 group and 105 objects from the CMa05, CMa07, and CMa08 groups. Consequently, 102 objects were rejected from CMa06, and 34 from CMa05, CMa07, and CMa08. As explained in Section \ref{subsec:subgroups}, we collectively refer to CMa05, CMa07, and CMa08 as one single population in the remainder of this paper. 

The discarded objects in the analysis predominantly exhibited probabilities between 50 \% and 80 \%. Our analysis uses a higher probability threshold compared to the \cite{Santos2021} paper where the standard threshold of 50\% is employed, which naturally leads to the rejection of some sources identified in that study. However, it is interesting to note that we have identified 1\,130 new members (see Fig. \ref{venn}). We have therefore tripled the number of cluster members associated with these subgroups of the CMa region with respect to the most recent census of the stellar population conducted with Gaia data \citep{Santos2021}.

\section{Analysis of radial velocities}
\label{subsection:radial}

The scarcity of radial velocity measurements is currently the main limitation to investigating the kinematic properties of the CMa region. The Gaia DR3 catalogue provides radial velocity information for 197 sources in our list of 1\,531 cluster members with a poor precision ranging from 2 to 40\,km\,s$^{-1}$ that is insufficient for many astrophysical purposes. We searched for high-resolution spectra in public databases, and we could not find any spectrum for our targets. We therefore resorted to making our own observations.

\begin{figure*}[!b]
\centering
\includegraphics[scale=0.49]{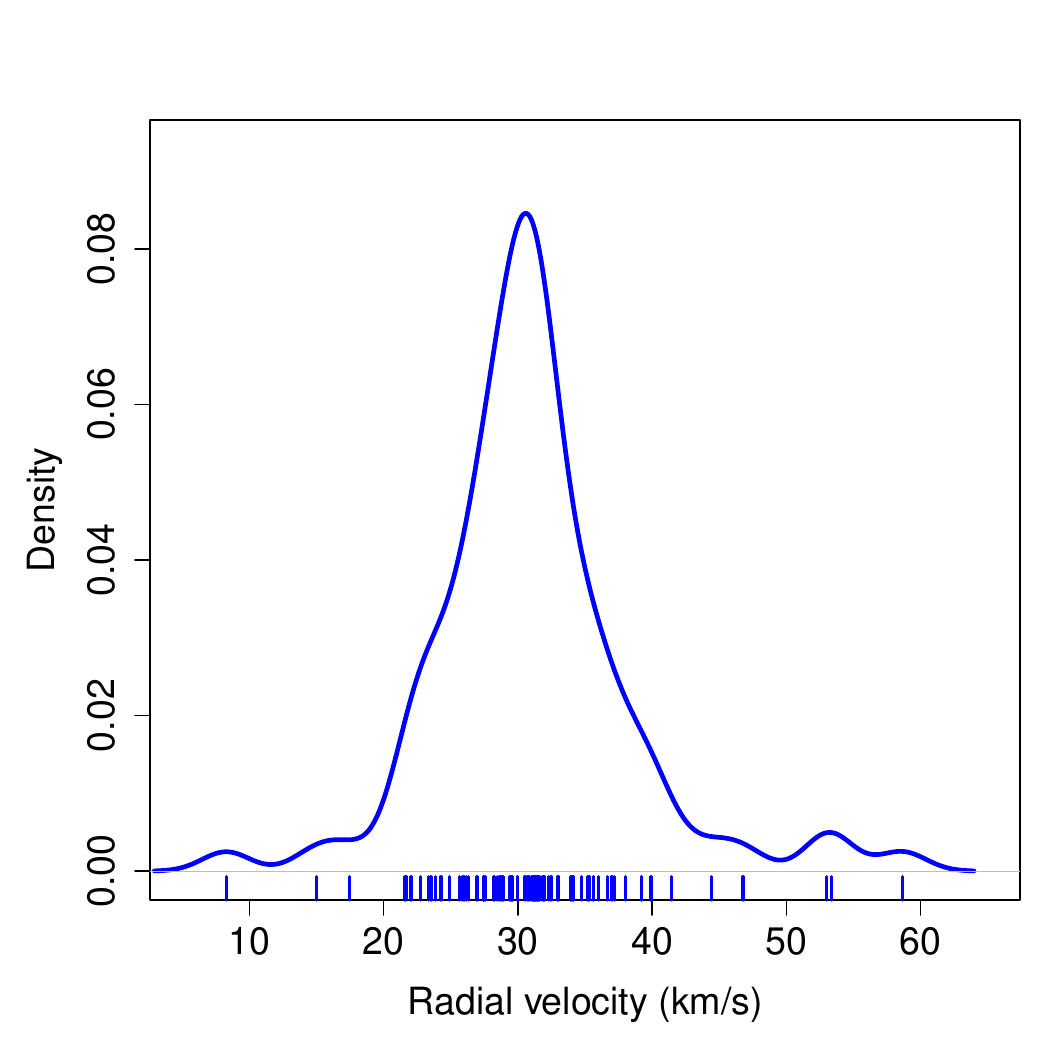}
\includegraphics[scale=0.49]{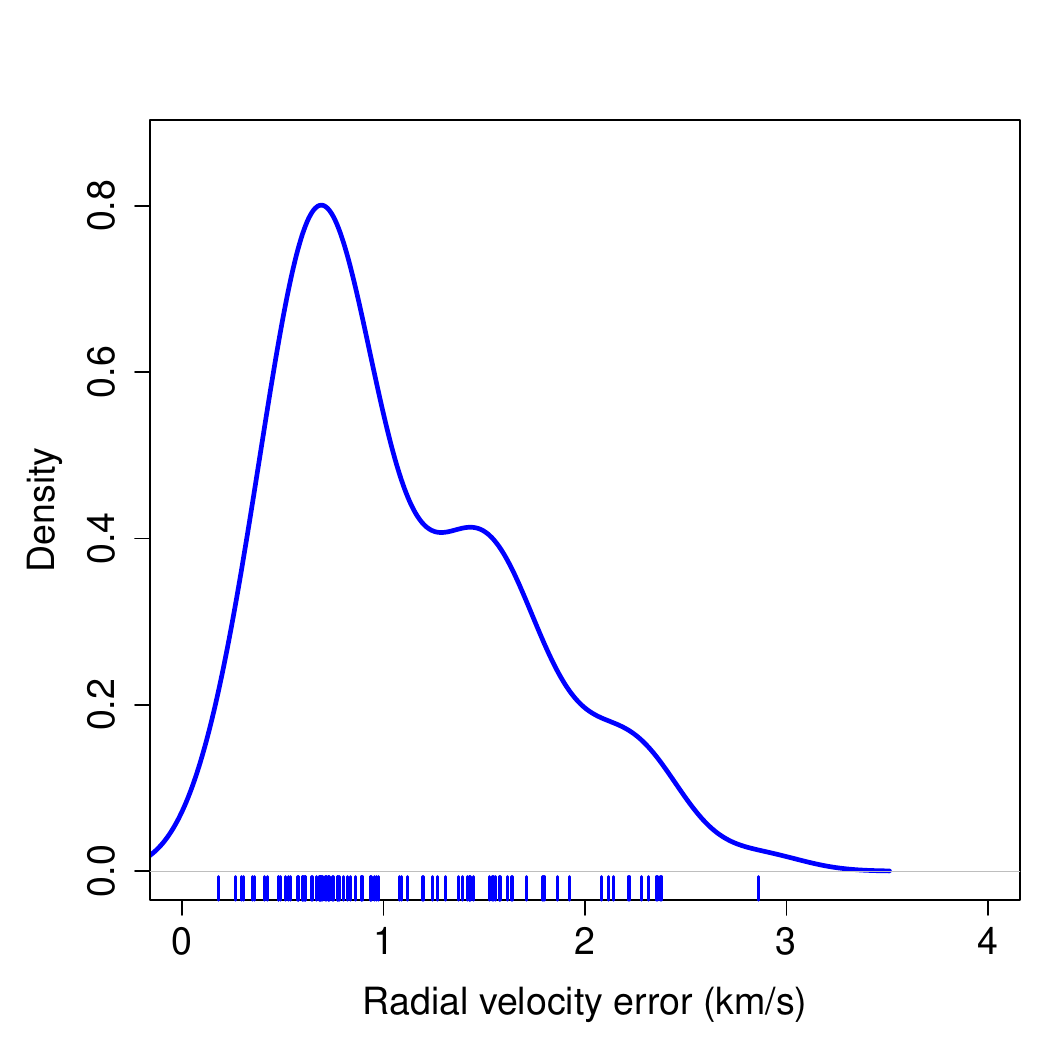}  
\caption{\textit{Left panel}: Kernel density estimation of the radial velocity distribution for the sample of 90 members with available measurements. \textit{Right panel}: Distribution of radial velocity uncertainties. The ticks in the horizontal axes mark the measurements for individual stars. }
 \label{four}
\end{figure*}

Our radial velocity analysis is based on observations (program 108.2250, PI: Galli) from the FLAMES spectrograph  \citep{flames}. The observations were conducted in service mode on the nights of November 12 and 16, 2021. FLAMES enables simultaneous observations with both UVES and GIRAFFE. For UVES, we utilized the 7+1 mode (580 nm setup, R=47000), employing seven fibres for target observations and one fibre to simultaneously record the ThAr lamp. For GIRAFFE, we used the MEDUSA mode (H665.0/HR15N setup, R=19000), with some fibres illuminated by a ThAr lamp for wavelength calibration. Most of our targets were observed with GIRAFFE, which can simultaneously observe up to 132 targets, including sky fibres. 
   
We observed six fields in the CMa region with 1800~s exposure times. To maximise fibre allocation, we targeted 700 sources across the region. However, only 188 were confirmed as cluster members and included in our analysis.

The spectra were reduced with the ESO pipeline, and afterwards we used iSpec \citep{blanco} to measure the radial velocity of each target. We determined radial velocities by cross-matching each spectrum with templates of different spectral types (A0, F0, G2, K0, K5, and M5) and selecting the one that has the closest spectral type to the target. We computed the cross-correlation function (CCF) for each pair of spectra (target-template) and visually inspected them to remove radial velocities that would result from a poor fit between the target spectrum and template, or a low signal-to-noise ratio S/N.

In doing so, we retained 90 high-quality spectra that produced reliable radial velocity measurements. The mean S/N for this sample was 48.6. The significant reduction in the number of spectra can be attributed to the characteristics of the data collection process. FLAMES operates as a multi-object spectrograph, capturing multiple sources simultaneously with the same exposure time. Consequently, faint sources would require longer exposures or individualized measurements. The lack of such adjustments for these sources resulted in spectra with low S/N and consequently poor CCFs that result in uncertain radial velocity measurements.

\begin{figure}[!h]
   \centering
   \includegraphics[scale=0.49]{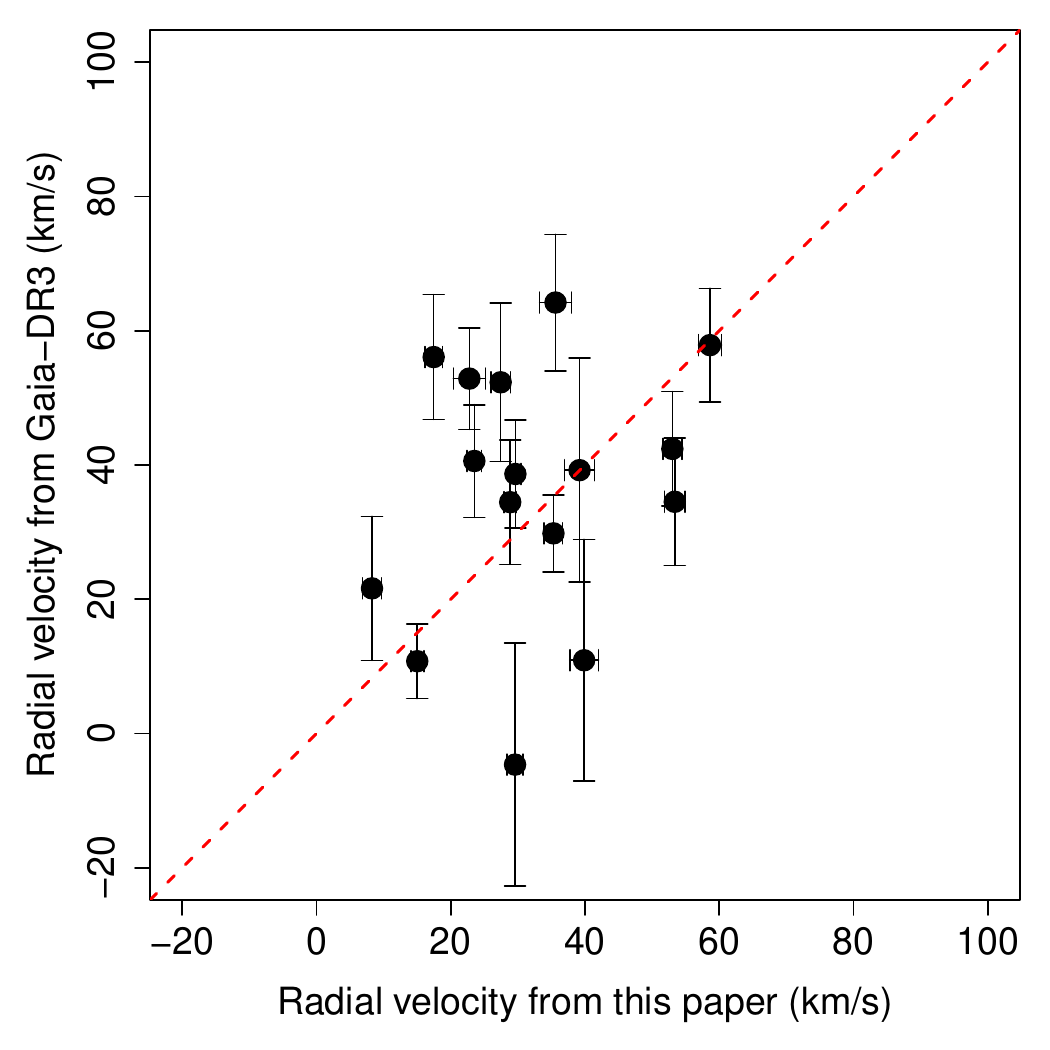}
   \caption{Comparison of radial velocity data obtained from our observations and Gaia DR3 data. One source of our sample (namely Gaia DR3 3046207139539172864) is not shown to improve the visibility of the plot.}
   \label{five}
\end{figure}

Fig. \ref{four} reveals an approximately symmetric radial velocity distribution with a peak near 30~\,km\,s$^{-1}$ and a few outliers (see Sect. \ref{subsec:spatial}). The uncertainty of individual measurements ranges from 0.2 to 2.9\,km\,s$^{-1}$, and the mean radial velocity uncertainty is about 1\,km\,s$^{-1}$. Fig. \ref{five} shows a comparison of the radial velocities derived in this paper with the ones given in the Gaia DR3 for 17 stars in common between the two projects. The mean difference between the radial velocities in the two datasets is 5~km\,s$^{-1}$, and the root mean square of the differences is about 42\,km\,s$^{-1}$. These high values arise from the poor precision of the radial velocities in Gaia DR3. The mean uncertainty of the Gaia DR3 radial velocity for this subsample is 10\,km\,s$^{-1}$. As explained above, our radial velocity measurements are more precise than those given in the Gaia DR3 catalogue. In the following we use these newly derived radial velocity data to investigate the kinematic properties of the CMa region.

\section{Properties of the CMa subgroups}\label{section4}

\subsection{Subgroups of the CMa region} \label{subsec:subgroups}
    
As illustrated in Fig. \ref{ravsdec}, our sample of CMa stars comprises multiple populations of young stars. In this section, we employ the partitioning around medoids \citep[PAM,][]{doi:https://doi.org/10.1002/9780470316801.ch2} clustering algorithm to explore the underlying structures and patterns within our sample of cluster members, aiming to deepen our understanding of the overall properties of the CMa region. 

Partitioning around medoids (PAM) is a robust clustering method that divides a dataset into $k$ distinct clusters. Unlike the $k$-means algorithm, which uses centroids (mean points) to represent clusters, PAM selects actual data points, called medoids, as cluster representatives. This approach makes PAM less sensitive to outliers and noise.
    
The algorithm begins by randomly selecting $k$ data points as initial medoids. Each data point is then assigned to the nearest medoid based on a specified distance metric, such as the Euclidean distance. Within each cluster, the medoid is iteratively updated to the most centrally located point, provided a better one exists. This assignment and update process continues until the medoids stabilize or a predefined number of iterations is reached. In this paper, we use the k-medoids routine from the PyClustering library \citep{pyclustering} that implements the PAM algorithm.
    
To determine the optimal number of clusters ($k$) for our dataset, we employed three widely used methods: the elbow method \citep{elbow}, calculated using pairwise distances from the \texttt{scikit-learn} library; the silhouette score \citep{silhouette}, also implemented in \texttt{scikit-learn}; and the gap statistic \citep{gap}, available in the \texttt{pyclustering} library. We performed the clustering analysis in a 3D space defined by proper motions and parallaxes. These methods suggest that the optimal number of clusters in our analysis is $k=2$.

\begin{figure*}[!htp]
  \centering
  \includegraphics[scale=0.4]{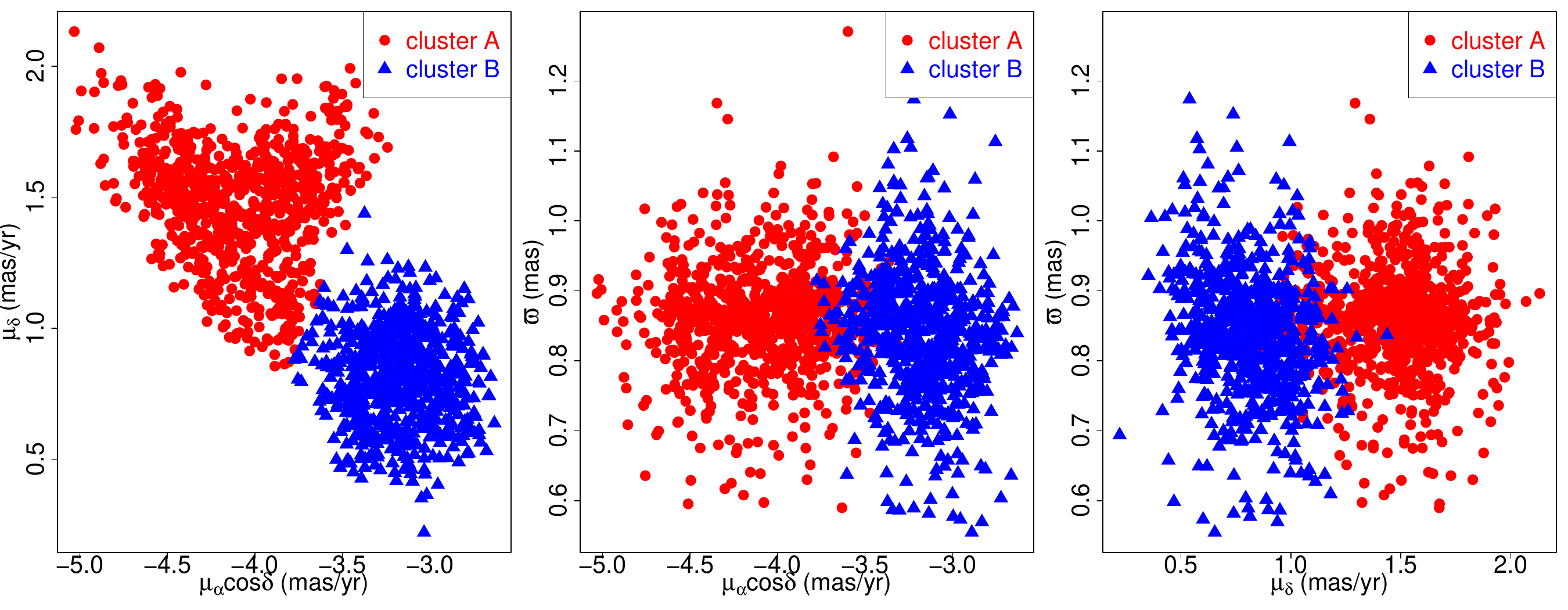}  
  \caption{PAM clustering results with $k=2$, showing identified groups.}
  \label{six}
\end{figure*}

Figure~\ref{six} displays the results of our clustering analysis performed with the PAM algorithm in the 3D astrometric space defined by proper motions ($\mu_{\alpha}\cos\delta$, $\mu_{\delta}$) and parallax ($\pi$). Two major subgroups were identified within CMa. The first subgroup, referred to as Cluster A, corresponds to the CMa06 subgroup defined by \cite{Santos2021}. The second subgroup, designated as Cluster B, encompasses sources previously assigned to CMa05, CMa07, and CMa08 in that study. Based on the higher-precision Gaia DR3 astrometry used in our analysis, we found no evidence of substructure within Cluster B. This outcome might be due to the specific clustering method applied (as \cite{Santos2021} used HDBSCAN) and the inclusion of newly identified sources in our sample. Therefore, throughout this paper, we collectively refer to CMa05, CMa07, and CMa08 as Cluster B.

The upcoming Gaia-DR4 catalogue, combined with ground-based spectroscopic surveys, will enable us in the near future to identify additional substructures within the CMa region, similar to what was done in Sco-Cen by \cite{2023A&A...677A..59R}, which could not be resolved in this study due to the relatively small number of sources with radial velocity measurements. This will provide new insights into the star formation history of CMa.

\subsection{Distance and spatial velocities}
\label{subsec:spatial}

We used our new list of cluster members identified in this study to re-visit the distance to CMa. We employed the Kalkayotl 2.0 method \citep{Olivares2025}, a Bayesian inference framework designed to infer the 3D positions and 3D velocities of stellar clusters (and individual members) based on Gaia DR3 astrometry. The  distances and velocities thus derived for individual sources in our sample are given in Table \ref{A3}.

We obtain the distance of $1\,165^{+95}_{-96}$~pc for the full sample of 1\,531 stars, which is consistent with values in the literature that place the region between 1\,000 and 1\,200 pc \citep{2008hsf2.book....1G}. This estimate is more precise than the distance of $1\,177^{+130}_{-107}$~pc calculated from the inverse of the mean parallax ($\varpi=0.850\pm0.085$~mas). Our result is also fully consistent with the mean of the photogeometric distance estimate of $1\,177^{+96}_{-121}$~pc that is computed from a direction-dependent prior on distance, the colour and apparent magnitude of the sources \citep{Bailer-Jones2021}. Moreover, we also note that the distances inferred for the two subgroups in our sample ($d_{A}=1\,150^{+79}_{-87}$~pc and $d_{B}=1\,183^{+103}_{-108}$~pc) are consistent between themselves and with the mean distance to CMa within the reported uncertainties (see Table \ref{tab_distance_velocity}).

\begin{figure*}[h]
\includegraphics[width=0.99\textwidth]{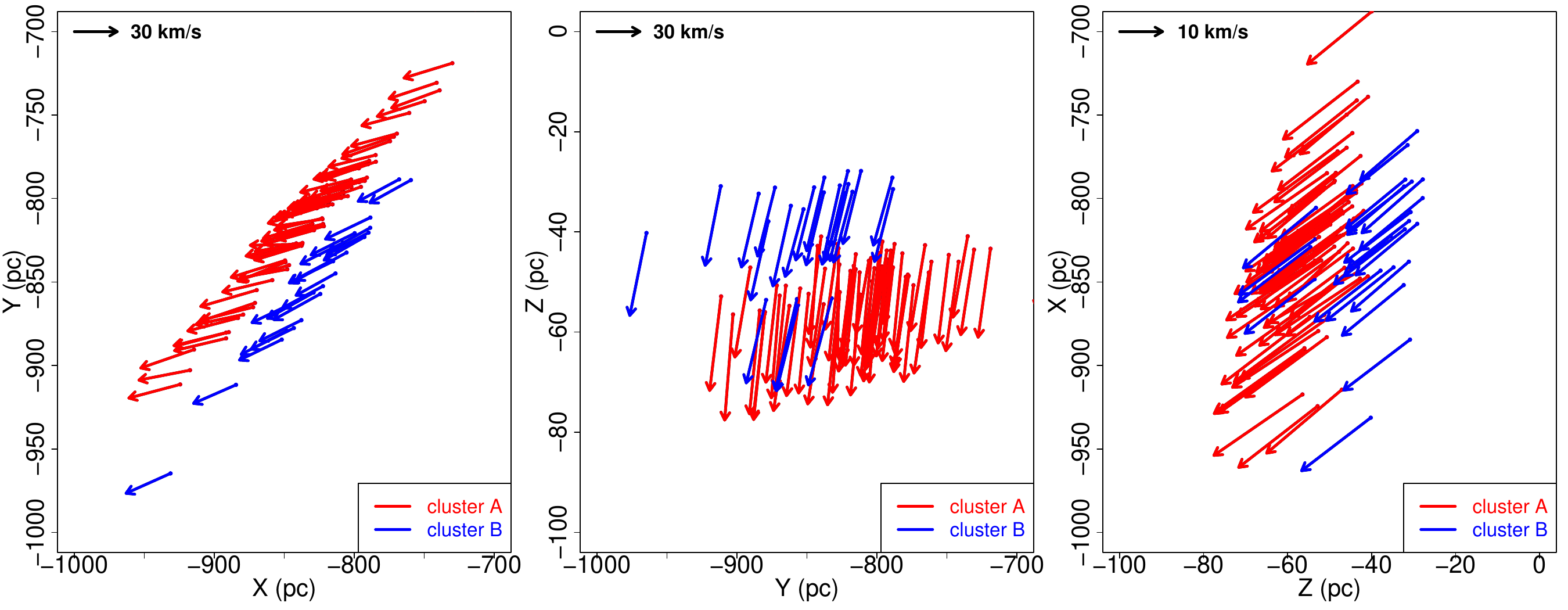}  
\caption{Spatial velocities of 82 members with known RVs projected on XY, YZ, and ZX planes.}
              \label{eight}%
    \end{figure*}

As mentioned before, Kakayotl 2.0 also returns the $UVW$ velocity components of the stars (see Table \ref{A3}). The $U$, $V$, and $W$ velocity components are given in a right-handed system with its origin at the Sun where $X$ points to the Galactic centre, $Y$ points to the direction of Galactic rotation, and $Z$ points to the Galactic North pole. We applied the interquartile range (IQR) criterion to identify and remove potential outliers from the distribution of the $U$,$V$, and $W$ velocity components. In doing so, we rejected eight stars (out of 90 stars, see Section \ref{subsection:radial}) yielding a final sample of 82 stars with complete data. Clusters A and B have 60 and 22 stars from this sample, respectively. Figure \ref{eight} presents the velocity vector of the stars. Although no significant differences are observed in the direction of motion between the two subgroups, there is a clear distinction in the spatial positions of the subgroups.

\begin{table*}[!h]
\centering
\footnotesize{
\caption{Distance and spatial velocity of CMa subgroups.
\label{tab_distance_velocity}}
\begin{tabular}{lcccccccccccc}
\hline\hline
Sample&$N_{d}$&$N_{UVW}$&$d$&\multicolumn{3}{c}{$U$}&\multicolumn{3}{c}{$V$}&\multicolumn{3}{c}{$W$}\\
&&&(pc)&\multicolumn{3}{c}{(km/s)}&\multicolumn{3}{c}{(km/s)}&\multicolumn{3}{c}{(km/s)}\\
\hline\hline
&&&&Mean&Median&SD&Mean&Median&SD&Mean&Median&SD\\
\hline
Cluster A & 828 & 60 & $1\,150^{+79}_{-88}$& $-34.0^{+1.3}_{-1.7}$ & -33.6 & 2.0 & $-7.7^{+1.6}_{-1.8}$ & -7.6 & 1.8 & $-18.3^{+1.7}_{-1.7}$& -18.4 & 1.7 \\
Cluster B & 703 & 22 & $1\,183^{+103}_{-108}$ &$-29.9^{+1.6}_{-1.6}$ & -29.5 & 1.9 & $-13.0^{+0.8}_{-0.8}$ & -12.7& 1.2 & $-14.8^{+1.4}_{-1.3}$ & -14.5 & 1.3 \\
\hline
CMa (all stars) & 1\,531 & 82 & $1\,165^{+95}_{-96}$ &$-32.9^{+3.1}_{-2.2}$& -33.1 & 2.7 & $-9.1^{+2.3}_{-3.4}$ & -8.5 & 2.9 & $-17.4^{+2.4}_{-2.2}$& -17.7 & 2.2 \\

\hline\hline
\end{tabular}
\tablefoot{We provide number of stars used to compute distance and spatial velocity, distance obtained from Bayesian inference, mean, median, and standard deviation (SD) of $UVW$ velocity components. The uncertainties in the distances and velocities are computed from the 16th and 84th percentiles of each distribution.}
}
\end{table*}

\begin{figure*}[!h]
\begin{center}
\includegraphics[width=0.99\textwidth]{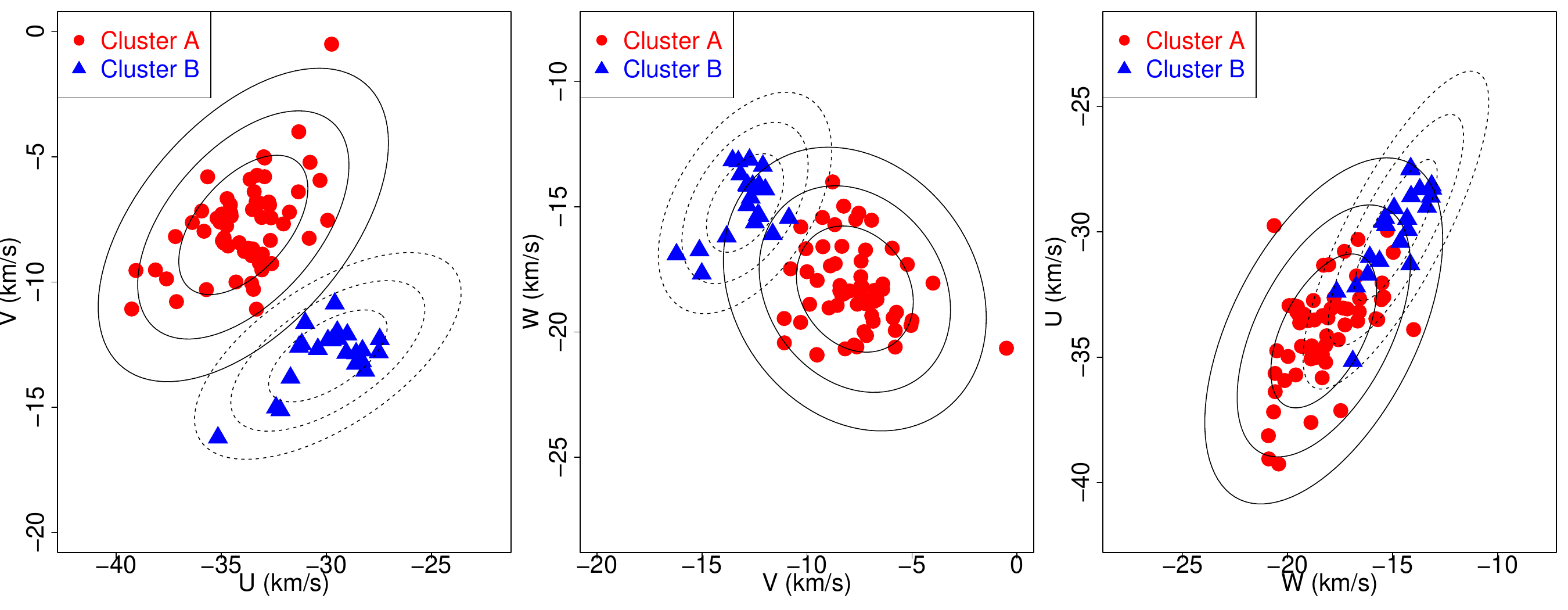}
\caption{Distribution of 3D velocity of CMa stars. The different symbols and colours indicate the two subgroups in our sample. The contours indicate the 68\%, 95.4\%, and 99.7\% confidence levels computed from the mean covariance matrix of each population. Solid and dashed lines indicate the confidence ellipses of Clusters A and B, respectively. \label{fig_UVW_ellipse} 
}
\end{center}
\end{figure*}

It is well known that the transformation of parallaxes, proper motions, and radial velocities into 3D velocities can result in correlated errors even in the absence of measurement errors \citep[see e.g.][]{Brown1997,Perryman1998}. Here, we employed the Kalkayotl 2.0 code that implements a simultaneous modelling of positions, velocities, and their correlations, and also parallax and proper motions angular correlations, making full use of the astrometric data given in the Gaia DR3 catalogue to derive accurate 3D positions and 3D velocities. Figure~\ref{fig_UVW_ellipse} illustrates the correlation among the velocity components as we compare the velocity distributions of the subgroups in our sample. There is a clear overlap between the velocity distribution of the two subgroups within the confidence region of 99.7\%, which suggests that their 3D space motions are still consistent. 

The median uncertainty in each component of the $UVW$ velocity of the full sample is 0.4, 0.3, and 0.3~km/s, respectively. These values are smaller than the observed velocity dispersion measured from the standard deviation in each direction (see Table~\ref{tab_distance_velocity}) suggesting that the intrinsic velocity dispersion of the CMa population is resolved ($\sigma_{U}^{int}\simeq2.6$~km/s, $\sigma_{V}^{int}\simeq2.9$~km/s, $\sigma_{W}^{int}\simeq2.2$~km/s). This implies that the velocity dispersion in CMa is somewhat consistent with the 1D velocity dispersion reported for other star-forming regions, for example Orion and Taurus, which typically range between 2 and 3 km/s \citep{Kounkel2018,Galli2019}.

\subsection{Ages}

\subsubsection{Isochronal ages}
\label{subsection:4.3.1}

In the following step we computed isochronal ages of the stars in our sample. The stellar ages were computed by interpolating the position of individual stars in the absolute colour-magnitude diagram constructed from Gaia DR3 photometry among the isochrones of pre-main sequence stars evolutionary models. We used three different models to compute isochronal ages: the \citet{baraffe} models (hereafter, BHAC15), PARSEC v.1.2S \citep{parsec}, and MIST \citep{mist}.

We computed isochronal ages for 1\,130, 1\,531, and 1\,531 cluster members using the BHAC15, PARSEC, and MIST models, respectively. Extinction corrections for the G band magnitudes were applied based on the Bayestar catalogue \citep{2016ApJ...821...78S}. BHAC15 isochrones do not cover the full range of colours displayed in our sample of CMa stars, which explains the smaller number of stars with ages computed from this model. 

Table~\ref{tab_ages} presents a comparative analysis of the stellar ages derived using three different evolutionary models. The mean isochronal age of the CMa region lies between approximately 2 and 3 Myr, depending on the adopted model. We do not find any significant age difference between the two subgroups, indicating that they are likely coeval. This result suggests that the CMa subgroups share a common formation history, similar to the stellar populations observed in the subgroups of the Lupus star-forming region \citep{Galli_2020}.

\begin{figure*}[!h]
\begin{center}
\includegraphics[width=0.47\textwidth]{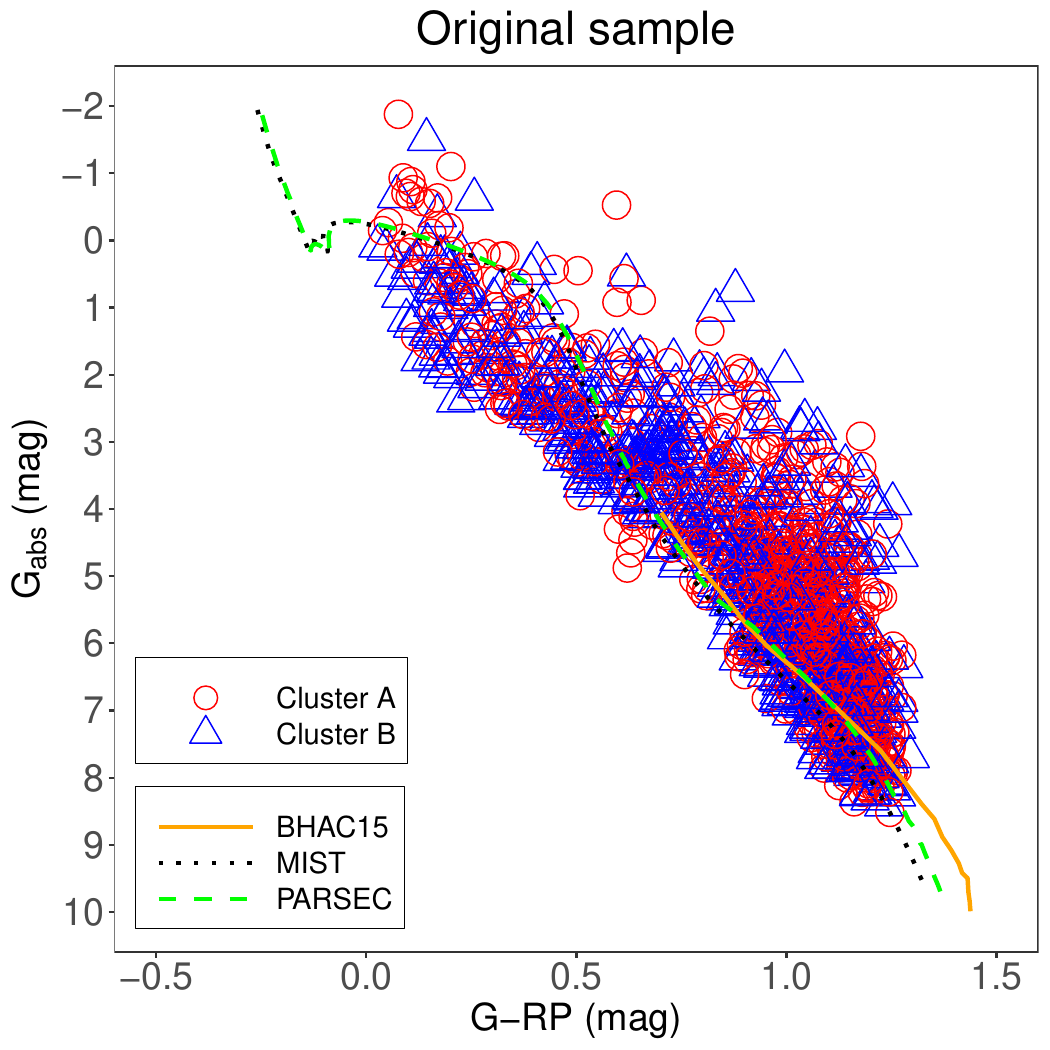}
\includegraphics[width=0.47\textwidth]{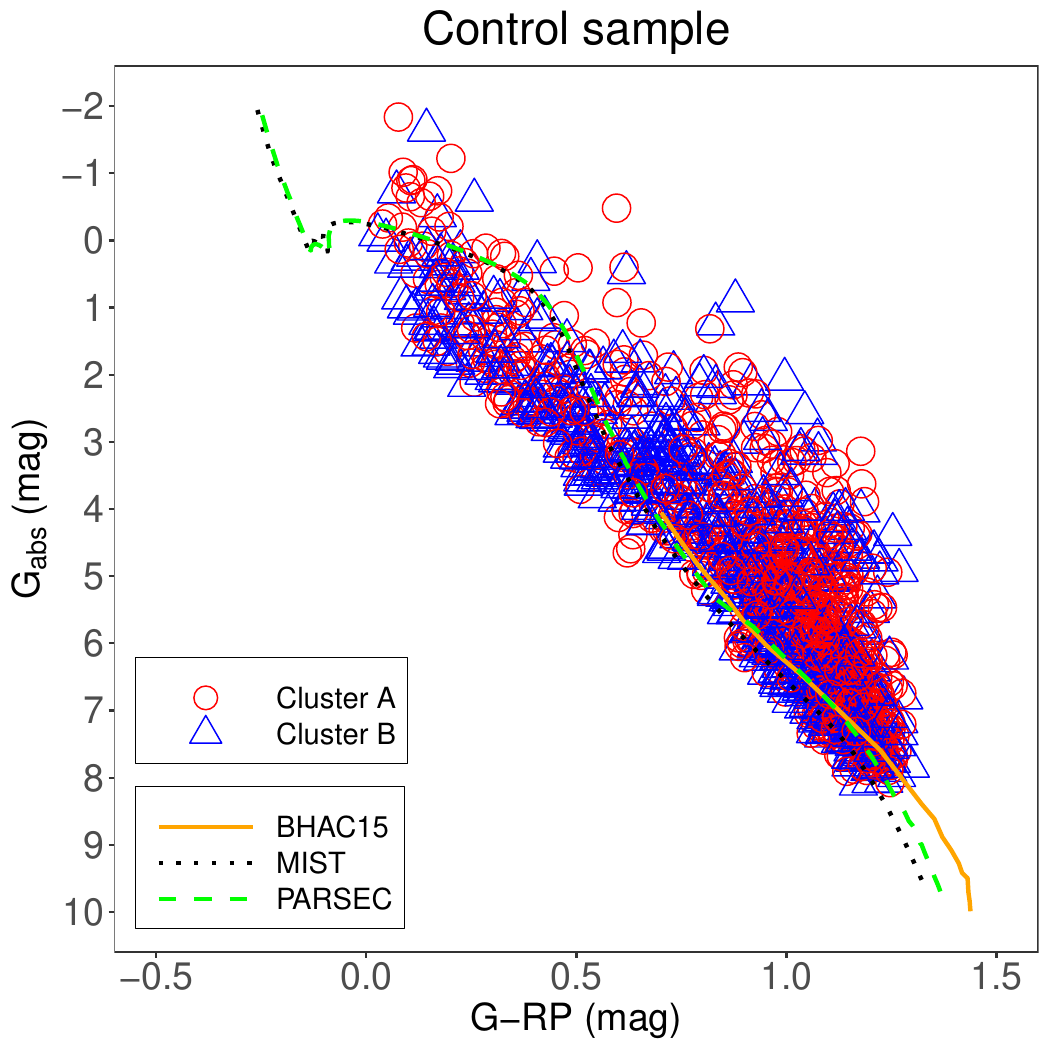}
\caption{Colour-magnitude diagram of CMa region obtained for our original sample \textit{(left panel)} and control sample \textit{(right panel)}. The different symbols indicate the two subgroups identified in our analysis, and the lines mark the 2~Myr isochrone obtained from different models.
\label{fig9} 
}
\end{center}
\end{figure*}

We constructed a control sample using the mean distance of the CMa region (see Table~\ref{tab_distance_velocity}) to compute the absolute magnitude for all stars. This was done to investigate whether the observed vertical spread in the colour-magnitude diagram can be explained by parallax measurement errors. However, as illustrated in Figure~\ref{fig9} we do not see significant changes in the colour-magnitude diagram caused by distance spread and the resulting ages are also fully consistent with those derived from our original sample (see Table~\ref{tab_ages}). This analysis confirms the robustness of our results and indicates that the stellar population under investigation is remarkably young, with some objects as young as approximately 1 Myr. In the Perseus region, \citet{perseus} reported a median stellar age of 10 Myr, while \citet{Bertout_2007} found that the Taurus region hosts a younger population, with a median age of approximately 5 Myr. In the case of Orion, \citet{Kounkel2018} identified subgroups with ages ranging between 2 and 7 Myr. Based on these results, we infer that the CMa sample is younger than both the Perseus and Taurus star-forming regions and more closely resembles the younger subgroups within Orion.

\begin{table*}[!h]
\centering
\footnotesize{
\caption{Isochronal age of CMa subgroups.
\label{tab_ages}}
\begin{tabular}{lccccccccccccccc}
\hline\hline
&\multicolumn{5}{c}{BHAC15}&\multicolumn{5}{c}{PARSEC}&\multicolumn{5}{c}{MIST}\\
\hline\hline
&&\multicolumn{2}{c}{Sample}&\multicolumn{2}{c}{Control}&&\multicolumn{2}{c}{Sample}&\multicolumn{2}{c}{Control}&&\multicolumn{2}{c}{Sample}&\multicolumn{2}{c}{Control}\\
&$N_{\star}$&Mean&SD&Mean&SD&$N_{\star}$&Mean&SD&Mean&SD&$N_{\star}$&Mean&SD&Mean&SD\\
&&(Myr)&(Myr)&(Myr)&(Myr)&&(Myr)&(Myr)&(Myr)&(Myr)&&(Myr)&(Myr)&(Myr)&(Myr)\\
\hline
Cluster A & 632 & 2.3  & 3.1  & 2.0 & 2.8 & 828 & 2.2 & 2.6 & 2.0 & 2.3 & 828 & 1.9 & 2.3 & 1.7 & 2.2 \\
Cluster B & 498 & 3.6 & 4.6 &  3.1 & 4.2 & 703 &  2.7 & 2.8 & 2.5 & 2.4 & 703 & 2.2 & 2.3 & 2.1 & 2.3\\
\hline
CMa (all stars) & 1\,130 & 2.9 & 3.9 & 2.5 & 3.6 & 1531 & 2.5  & 2.7 &  2.2 & 2.4& 1531 & 2.0 & 2.3& 1.9 & 2.2  \\

\hline\hline
\end{tabular}
\tablefoot{ For each sample number of stars, we provide the mean and standard deviation (SD) of the ages derived from the different models. Age results are presented for the original sample (using individual distance estimates for each star) and control sample (using the mean distance of the CMa region for all stars).}
}
\end{table*}

\subsubsection{ Fraction of disc-bearing stars}
\label{subsection:class}

\begin{table*}[b]
\centering
\caption{Results of classification scheme of \cite{Koenig_2014} applied to our sample.}
\small
\resizebox{0.69\textwidth}{!}{
\begin{tabular}{c c c c c c} 
\hline\hline
Cluster & Members& Class I & Class II & Class III & not classified \\ 
\hline\hline
A       & 828 & 0     & 44  & 110 & 674 \\
B       & 703 & 0     & 23  & 250 & 430\\
Total   & 1\,531 & 0     & 67  & 360 & 1\,104 \\ 
\hline
\end{tabular}}
\label{4}
\end{table*}

To complement our age analysis, we assessed the fraction of disc-bearing stars in our sample of cluster members. In this context, we used the methodology developed by \cite{Koenig_2014} to detect and classify young stellar objects based on their infrared excess emission. We cross-matched our list of CMa members with the 2MASS and AllWISE catalogues to retrieve the infrared photometry of the stars in our sample. To do so, we used the pre-computed cross-match tables available in the Gaia archive with both the 2MASS and AllWISE catalogues. We found 2MASS and AllWISE photometry for 1\,031 stars (in our sample of 1\,531 cluster members), but 604 of them exhibit poor measurements that were discarded according to the \cite{Koenig_2014} quality criteria. This leaves us with an effective sample of 427 sources with available infrared photometry for this analysis. 

\begin{figure}[h]
   \centering
     \includegraphics[scale=0.34]{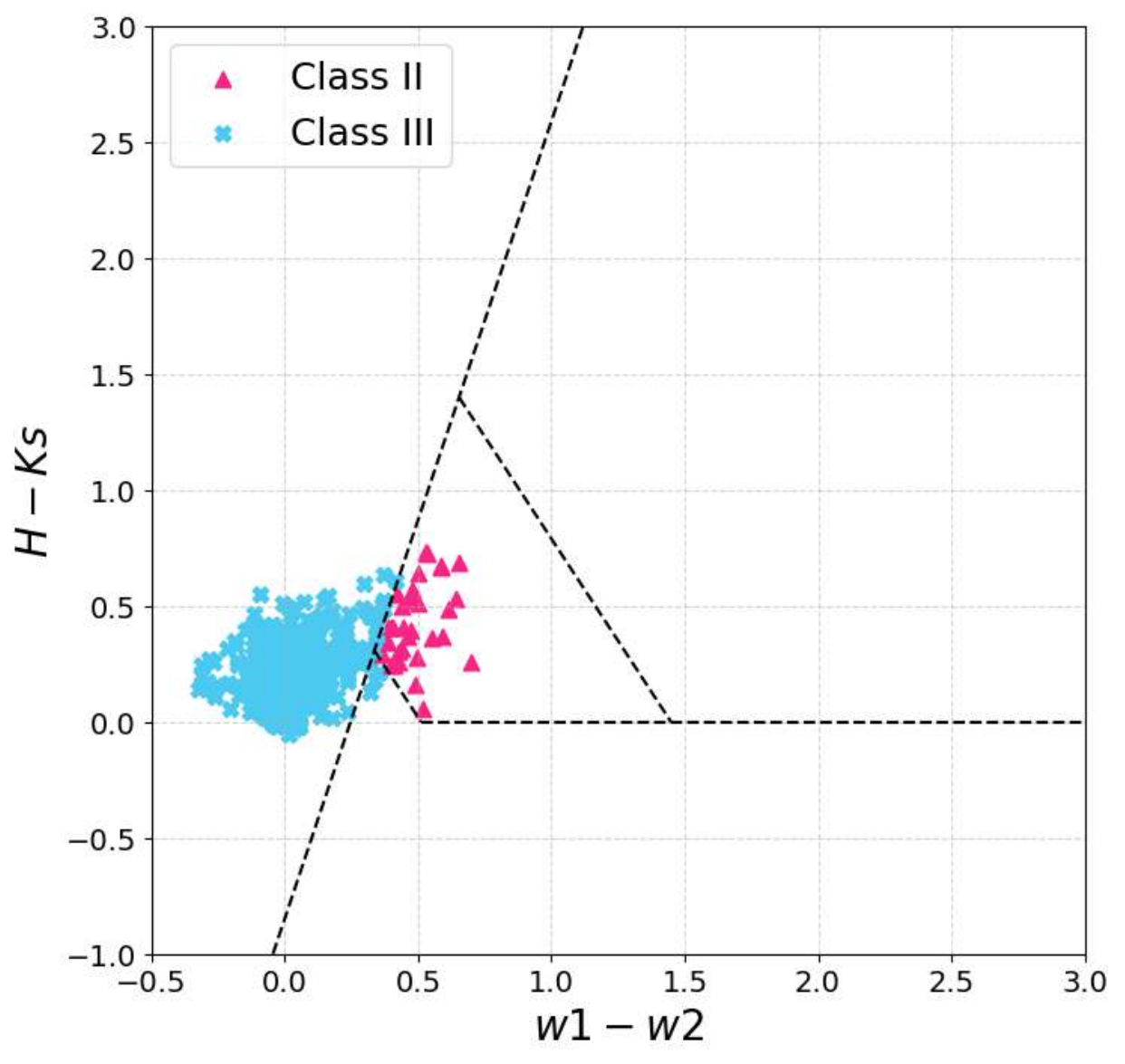}
     \includegraphics[scale=0.34]{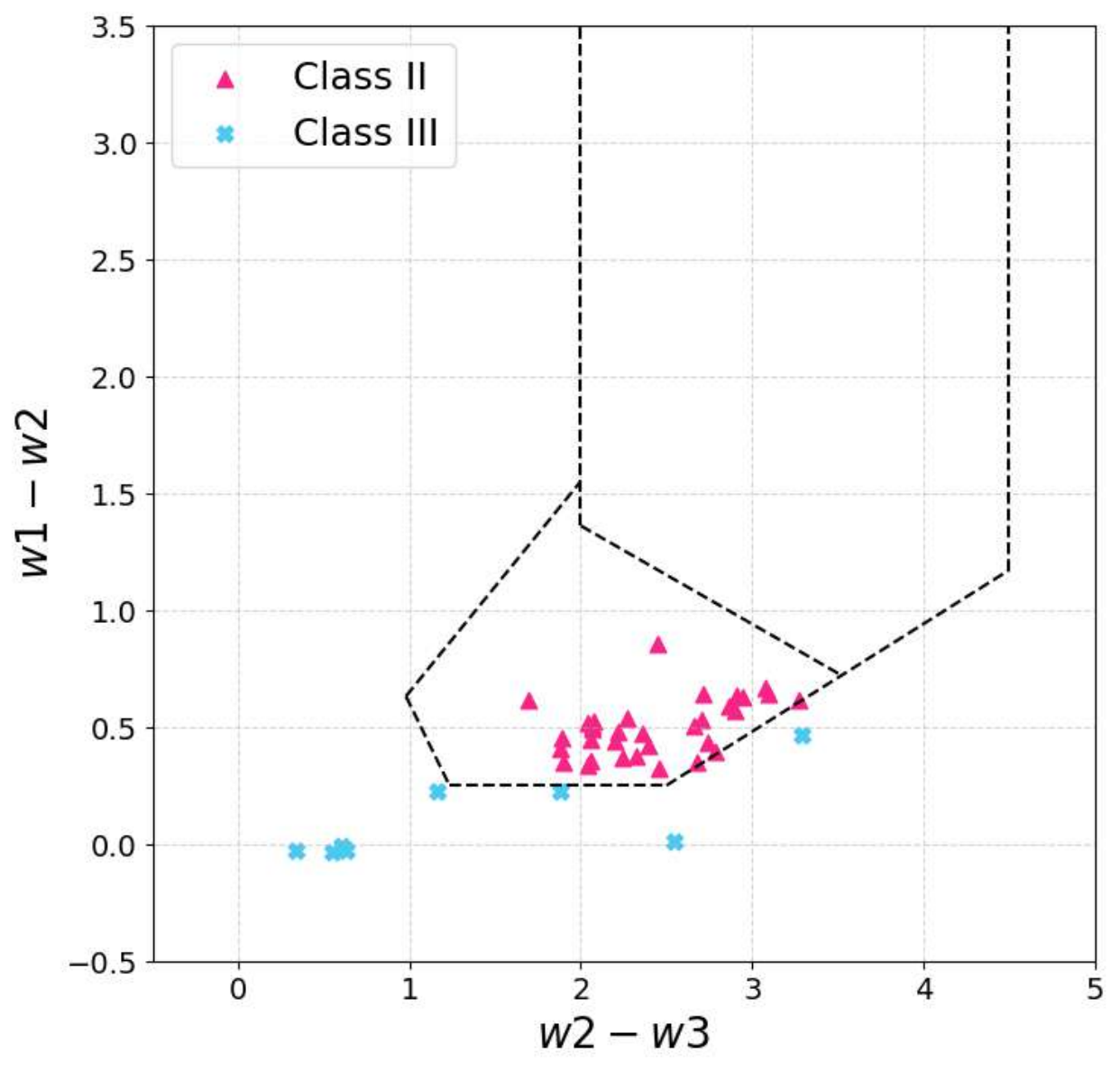}
   \caption{ Classification of objects using method developed by \cite{Koenig_2014}.}
              \label{ten}
    \end{figure}

In our sample, we identified 67 Class II stars based on the classification scheme proposed by \citet{Koenig_2014}. No Class I sources were detected. One object, Gaia DR3 3048979420667363072, was initially classified as an active galactic nucleus (AGN); however, its estimated age (2 Myr; see Sect.~\ref{subsection:4.3.1}) and parallax ($\varpi = 0.918 \pm 0.062$ mas) are inconsistent with this classification. We therefore consider this source to have been misclassified. The remaining 359 stars were not assigned a class by this method; nevertheless, considering their young ages (see Fig.~\ref{fig9}) and their loci in the colour–colour diagrams (Fig.~\ref{ten}), it is plausible to classify them as Class III objects. A similar reasoning applies to Gaia DR3 3048979420667363072, bringing the total number of Class III sources to 360.
    
In Table \ref{4}, we present the number of objects corresponding to each subclass of  YSOs identified in the two clusters of our sample. In Cluster A, 44 members were classified as Class II (28.6\%), while 110 were identified as Class~III (71.4\%). For Cluster B, a total of 273 sources were classified, comprising 23 Class II stars (8.4\%) and 249 Class III stars (91.2\%). When comparing the fraction of disc-bearing stars between the two populations, we observe a slight indication that Cluster B exhibits a lower disc fraction. The separation between the classifications is illustrated in Fig. \ref{ten}.

When comparing the proportion of classified sources within each subgroup, there is a subtle indication that Cluster B may be at a more advanced evolutionary stage than Cluster A, suggesting that the latter could be relatively younger. It is interesting to note that Cluster~A, which appears to be somewhat younger (see Table~\ref{tab_ages}), is also more concentrated in the vicinity of the molecular clouds of the CMa region while Cluster~B defines a more dispersed population of young stars (see Figure~\ref{ravsdec}). However, the limited number of classified sources in this region prevents us from drawing firm conclusions. The most plausible scenario, based on our results, is that both clusters formed contemporaneously.

\section{Conclusions}\label{section5}

In this paper, we performed a new membership analysis of the CMa region based on Gaia DR3 data. We inferred membership probabilities for 162\,331 sources in the field of CMa and identified 1\,531 high-probability cluster members. We confirm 401 members that were known from previous studies and identified 1\,130 new members. We have therefore increased the number of cluster members in this region by a factor of 3 with respect to the previous study in this region by \cite{Santos2021}.

We successfully calculated radial velocities for 90 sources in the CMa region using observations conducted by our team with the FLAMES spectrograph. The radial velocities derived in this paper have a typical (mean) uncertainty of about 1\,km\,s$^{-1}$. This represents a major improvement with respect to the Gaia DR3 catalogue where the mean radial velocity uncertainty of the stars in our sample is about 10~\,km\,s$^{-1}$. This is the most precise radial velocity survey of the CMa region to date.

We used our new census of cluster members to investigate the structure of the CMa region using the PAM clustering and identified two subgroups labelled Cluster A and Cluster B. The former corresponds to the CMa06 subgroup and the latter encompasses CMa05, CMa07, and CMa08, which were all previously identified by \cite{Santos2021}. We estimate the distance of the two subgroups in CMa from Bayesian inference and show that they are roughly located at the same distance ($d= 1\,165^{+95}_{-96}$~pc) and have similar space motions.

We investigated the age of the CMa subgroups using two different approaches. First, we computed isochronal ages from different models, and we showed that the two subgroups have a mean age of 2-3 Myr. We do not see significant age differences between the two subgroups with this method. Second, we investigated the fraction of disc-bearing stars in the sample based on their infrared excess emission. We show that the fraction of disc-bearing stars in Cluster A is somewhat higher, implying that it may be at a younger evolutionary stage. However, the high number of sources with missing data for this analysis prevents us from drawing firm conclusions and a more careful analysis will be necessary in future studies.

The main limitation to this study was the lack of high-resolution spectroscopy, which is crucial for characterizing the cluster members individually in terms of their physical properties and kinematics. In particular, radial velocities are the missing piece for unlocking the 3D space motion of the stars and reconstructing the local history of star formation of the CMa region. We therefore encourage astronomers to perform high-resolution spectroscopy of the CMa region, which combined with the precise astrometry delivered by the Gaia satellite will allow for a much more complete picture of this star-forming complex.

\section*{Data availability}

Tables A.1, A.2, and A.3 are only available in electronic form at the CDS via anonymous ftp to cdsarc.u-strasbg.fr (130.79.128.5) or via http://cdsweb.u-strasbg.fr/cgi-bin/qcat?J/A+A/.

\begin{acknowledgements} 
We thank the referee for constructive criticism. S.N.S acknowledges financial support from the São Paulo Research Foundation (FAPESP, grant: 2022/06054-4). P.A.B.G. acknowledges financial support from the São Paulo Research Foundation (FAPESP, grant: 2020/12518-8) and Conselho Nacional de Desenvolvimento Científico e Tecnológico (CNPq, grant: 303659/2024-6). JO acknowledge financial support from: "Ayudas para contratos postdoctorales de investigación UNED 2021" and project PID2022-142707NA-I00 financed by MCIN/AEI/10.13039/501100011033/FEDER, UE. NMR acknowledges support from the Beatriu de Pinos' postdoctoral program under the Ministry of Reseach and Universities of the Government of Catalonia (Grant Reference No. 2023 BP 00215). J.G.H. acknowledges financial support from the São Paulo Research Foundation (FAPESP, grant: 2023/08726-2). This work has made use of data from the European Space Agency (ESA) mission Gaia (https://www.cosmos.esa.int/gaia), processed by the Gaia Data Processing and Analysis Consortium (DPAC, https://www.cosmos.esa.int/web/gaia/dpac/consortium). This research has also made use of data products from the Two Micron All Sky Survey (2MASS) and the AllWISE program, which combines data from the WISE and NEOWISE missions. Additionally, this study has made use of the SIMBAD database, operated at CDS, Strasbourg, France.
\end{acknowledgements} 

\bibliographystyle{aa}
\bibliography{references}

\begin{appendix}
\section{Tables (online material)}\label{appendix_tables}

\begin{table*}[!h]
\centering
\caption{Properties of 1531 cluster members selected from our membership analysis.
\label{A1}}
\scriptsize{
\begin{tabular}{cccccccccc}
\hline\hline
Designation&$\alpha$&$\delta$&$\mu_{\alpha}\cos\delta$&$\mu_{\delta}$&$\varpi$&G&G-RP&probability&Cluster\\
&(deg) & (deg)&(mas/yr)&(mas/yr)&(mas)&(mag)&(mag)\\
\hline\hline

Gaia DR3 3048968764848107008 & 104.011186 & -11.54237 & $ -4.510 \pm 0.158 $& $ 1.676 \pm 0.178 $& $ 0.595 \pm 0.156 $& 18.346 & 1.140 & 0.8927 & A \\
Gaia DR3 3049464717612162432 & 104.062796 & -10.19907 & $ -3.285 \pm 0.039 $& $ 0.644 \pm 0.038 $& $ 0.730 \pm 0.041 $& 15.857 & 0.943 & 0.8641 & B \\
Gaia DR3 3048945404526512128 & 104.066009 & -11.75762 & $ -4.370 \pm 0.041 $& $ 1.451 \pm 0.049 $& $ 0.881 \pm 0.042 $& 15.958 & 0.817 & 0.9963 & A \\
Gaia DR3 3048965023937085056 & 104.079029 & -11.62532 & $ -3.269 \pm 0.023 $& $ 1.167 \pm 0.024 $& $ 0.765 \pm 0.020 $& 14.416 & 0.737 & 0.8690 & B \\
Gaia DR3 3049366517480046976 & 104.114280 & -10.58022 & $ -3.331 \pm 0.035 $& $ 0.938 \pm 0.036 $& $ 0.766 \pm 0.035 $& 15.460 & 0.682 & 0.8942 & B \\
Gaia DR3 2949529143692209152 & 104.131291 & -13.25642 & $ -2.955 \pm 0.028 $& $ 0.723 \pm 0.031 $& $ 0.856 \pm 0.028 $& 15.033 & 0.597 & 0.9124 & B \\
Gaia DR3 3045933081963199616 & 104.149088 & -12.09717 & $ -4.388 \pm 0.074 $& $ 1.471 \pm 0.073 $& $ 0.759 \pm 0.074 $& 16.884 & 0.961 & 0.9742 & A \\
Gaia DR3 3049045528803431040 & 104.165136 & -11.27261 & $ -3.561 \pm 0.187 $& $ 1.366 \pm 0.218 $& $ 0.839 \pm 0.185 $& 18.393 & 1.136 & 0.8616 & A \\
Gaia DR3 3049413414224664192 & 104.172107 & -10.26195 & $ -3.548 \pm 0.066 $& $ 0.578 \pm 0.057 $& $ 0.912 \pm 0.061 $& 16.637 & 1.024 & 0.9345 & B \\
Gaia DR3 3048994985628364160 & 104.181811 & -11.41184 & $ -3.221 \pm 0.082 $& $ 0.714 \pm 0.095 $& $ 0.973 \pm 0.084 $& 17.096 & 1.131 & 0.8601 & B \\
Gaia DR3 3049045700602080512 & 104.204650 & -11.23772 & $ -4.078 \pm 0.202 $& $ 1.324 \pm 0.210 $& $ 0.597 \pm 0.194 $& 14.912 & 0.721 & 0.9656 & A \\
Gaia DR3 3048953891381829248 & 104.212736 & -11.62527 & $ -3.464 \pm 0.070 $& $ 0.724 \pm 0.092 $& $ 0.920 \pm 0.074 $& 16.954 & 0.927 & 0.9130 & B \\
Gaia DR3 3049173896787544576 & 104.242977 & -10.71995 & $ -3.686 \pm 0.124 $& $ 1.217 \pm 0.113 $& $ 0.703 \pm 0.116 $& 14.926 & 0.744 & 0.9226 & A \\
Gaia DR3 3048954441137629312 & 104.246435 & -11.58722 & $ -2.746 \pm 0.026 $& $ 0.601 \pm 0.033 $& $ 0.835 \pm 0.025 $& 14.869 & 0.724 & 0.9507 & B \\
Gaia DR3 3049400155663764224 & 104.276029 & -10.34730 & $ -3.001 \pm 0.085 $& $ 0.585 \pm 0.076 $& $ 0.828 \pm 0.082 $& 17.312 & 1.122 & 0.9193 & B \\
Gaia DR3 3048997283431523584 & 104.280136 & -11.33100 & $ -3.222 \pm 0.114 $& $ 0.806 \pm 0.118 $& $ 0.589 \pm 0.119 $& 17.971 & 1.101 & 0.9046 & B \\
Gaia DR3 3048949046658735360 & 104.280823 & -11.74521 & $ -4.731 \pm 0.051 $& $ 1.803 \pm 0.062 $& $ 0.948 \pm 0.047 $& 15.940 & 0.796 & 0.9666 & A \\
Gaia DR3 3049368544704861824 & 104.301023 & -10.59597 & $ -4.034 \pm 0.033 $& $ 1.417 \pm 0.028 $& $ 0.881 \pm 0.030 $& 10.887 & 0.039 & 0.9803 & A \\
Gaia DR3 3048973575214008704 & 104.334831 & -11.64085 & $ -3.394 \pm 0.018 $& $ 0.502 \pm 0.021 $& $ 0.929 \pm 0.018 $& 12.259 & 0.154 & 0.9751 & B \\
Gaia DR3 3049149432653966336 & 104.356105 & -11.04231 & $ -3.612 \pm 0.050 $& $ 0.765 \pm 0.044 $& $ 0.960 \pm 0.046 $& 16.285 & 0.871 & 0.9286 & B \\

\hline\hline
\end{tabular}
}
\tablefoot{We provide for each source Gaia DR3 identifier, position, proper motion, parallax, G-band photometry, G-RP colour, membership probability (scaled from zero to one), and subgroup of the CMa region. This table will be available in its entirety in machine-readable form at the CDS.}
\end{table*}

\begin{table*}[ht!]
\centering
\caption{Membership probability of 162\,331 sources in the input catalogue.}
\label{A2}
\begin{tabular}{lcccc}
\hline
Designation & probability & probability & probability & probability \\
 & {$p_{in}$ = 0.6} & {$p_{in}$ = 0.7} & {$p_{in}$ = 0.8} & {$p_{in}$ = 0.9} \\
\hline
Gaia DR3 3049062910532370304 & $3.65 \times 10^{-39}$ & $5.00 \times 10^{-41}$ & $2.80 \times 10^{-53}$ & $2.02 \times 10^{-93}$ \\
Gaia DR3 3048968562989718272 & $3.92 \times 10^{-13}$ & $1.35 \times 10^{-13}$ & $5.41 \times 10^{-16}$ & $7.89 \times 10^{-22}$ \\
Gaia DR3 3049390874236126720 & $2.99 \times 10^{-22}$ & $1.76 \times 10^{-41}$ & $4.63 \times 10^{-46}$ & $1.21 \times 10^{-99}$ \\
Gaia DR3 2949434826209329152 & $9.56 \times 10^{-25}$ & $7.62 \times 10^{-26}$ & $1.64 \times 10^{-34}$ & $6.12 \times 10^{-61}$ \\
Gaia DR3 3049367337814950656 & $5.65 \times 10^{-88}$ & $5.39 \times 10^{-165}$ & $6.01 \times 10^{-193}$ & $2.88 \times 10^{-116}$ \\
Gaia DR3 3049264847014583808 & $2.42 \times 10^{-38}$ & $5.11 \times 10^{-44}$ & $1.36 \times 10^{-62}$ & $1.98 \times 10^{-136}$ \\
Gaia DR3 3049470902364600960 & $1.95 \times 10^{-294}$ & $1.95 \times 10^{-294}$ & $1.93 \times 10^{-294}$ & $1.92 \times 10^{-294}$ \\
Gaia DR3 2949748423242713472 & $4.50 \times 10^{-22}$ & $5.93 \times 10^{-26}$ & $3.09 \times 10^{-31}$ & $2.17 \times 10^{-31}$ \\
Gaia DR3 2949824053321627136 & $1.36 \times 10^{-293}$ & $1.36 \times 10^{-293}$ & $1.35 \times 10^{-293}$ & $1.34 \times 10^{-293}$ \\
Gaia DR3 2949856935595134080 & $3.30 \times 10^{-10}$ & $1.95 \times 10^{-14}$ & $4.28 \times 10^{-19}$ & $9.48 \times 10^{-31}$ \\
Gaia DR3 3049062532576337280 & $2.57 \times 10^{-5}$ & $3.06 \times 10^{-4}$ & $1.06 \times 10^{-6}$ & $3.64 \times 10^{-16}$ \\
Gaia DR3 2949623285078111616 & $5.46 \times 10^{-22}$ & $2.82 \times 10^{-47}$ & $1.53 \times 10^{-51}$ & $7.71 \times 10^{-144}$ \\
Gaia DR3 3048956915038879360 & $2.52 \times 10^{-68}$ & $1.20 \times 10^{-69}$ & $7.85 \times 10^{-98}$ & $2.41 \times 10^{-152}$ \\
Gaia DR3 2949854216875845888 & $4.28 \times 10^{-83}$ & $8.15 \times 10^{-176}$ & $3.67 \times 10^{-223}$ & $7.85 \times 10^{-298}$ \\
Gaia DR3 2949428126060642816 & $1.03 \times 10^{-9}$ & $4.06 \times 10^{-19}$ & $6.10 \times 10^{-22}$ & $1.12 \times 10^{-65}$ \\
Gaia DR3 2949748732482404352 & $4.12 \times 10^{-45}$ & $3.04 \times 10^{-54}$ & $6.09 \times 10^{-63}$ & $1.12 \times 10^{-95}$ \\
Gaia DR3 3048945026569411584 & $2.15 \times 10^{-40}$ & $2.84 \times 10^{-56}$ & $8.05 \times 10^{-63}$ & $1.15 \times 10^{-90}$ \\
Gaia DR3 2949623323735795968 & $1.12 \times 10^{-7}$ & $4.12 \times 10^{-9}$ & $6.09 \times 10^{-11}$ & $1.28 \times 10^{-17}$ \\
Gaia DR3 3049386721006185472 & $1.36 \times 10^{-22}$ & $3.84 \times 10^{-27}$ & $1.21 \times 10^{-32}$ & $2.39 \times 10^{-43}$ \\
Gaia DR3 2949622705260528128 & $2.85 \times 10^{-261}$ & $2.10 \times 10^{-296}$ & $2.09 \times 10^{-296}$ & $2.07 \times 10^{-296}$ \\
\hline
\end{tabular}
\tablefoot{We provide for each source  membership probability computed for various analyses conducted in this study using different values for $p_{in}$ probability threshold. This table will be available in its entirety in machine-readable form at the CDS.}
\end{table*}

\begin{landscape}
\begin{table}
\centering
\caption{Properties of 90 members with radial velocities measured in our analysis.
\label{A3}}
\scriptsize{
\begin{tabular}{ccccccccccccc}
\hline\hline
Designation&$\alpha$&$\delta$&$\mu_{\alpha}\cos\delta$&$\mu_{\delta}$&$\varpi$&RV&distance&U&V&W&Cluster&flag\\
&(deg) & (deg)&(mas/yr)&(mas/yr)&(mas)&(km/s)&(pc)&(km/s)&(km/s)&(km/s)\\
\hline\hline

Gaia DR3 3046024586247581312 & 105.802880 & -11.47730 & $ -3.713 \pm 0.039 $& $ 1.420 \pm 0.038 $& $ 0.882 \pm 0.040 $& $ 22.0 \pm 1.3 $& $ 1113.4 ^{+ 0.8 }_{ -0.8 } $& $ -29.9 ^{+ 0.6 }_{ -0.6 } $& $ -7.5 ^{+ 0.5 }_{ -0.6 } $& $ -15.3 ^{+ 0.4 }_{ -0.4 } $& A & 1 \\
Gaia DR3 3046020016402397184 & 105.858956 & -11.58402 & $ -4.611 \pm 0.111 $& $ 1.563 \pm 0.114 $& $ 0.891 \pm 0.111 $& $ 28.7 \pm 0.4 $& $ 1103.4 ^{+ 2.3 }_{ -2.3 } $& $ -32.7 ^{+ 1.0 }_{ -0.9 } $& $ -6.8 ^{+ 1.0 }_{ -0.9 } $& $ -18.8 ^{+ 1.1 }_{ -1.1 } $& A & 1 \\
Gaia DR3 3046020841036099328 & 105.876379 & -11.53413 & $ -3.676 \pm 0.074 $& $ 1.500 \pm 0.069 $& $ 0.862 \pm 0.072 $& $ 32.0 \pm 2.9 $& $ 1139.2 ^{+ 1.9 }_{ -1.7 } $& $ -32.6 ^{+ 1.1 }_{ -1.0 } $& $ -9.3 ^{+ 1.0 }_{ -1.0 } $& $ -15.4 ^{+ 0.8 }_{ -0.9 } $& A & 1 \\
Gaia DR3 3046019226128411008 & 105.886981 & -11.59314 & $ -4.086 \pm 0.029 $& $ 0.984 \pm 0.026 $& $ 0.820 \pm 0.027 $& $ 36.7 \pm 1.6 $& $ 1194.4 ^{+ 0.7 }_{ -0.7 } $& $ -33.3 ^{+ 0.4 }_{ -0.4 } $& $ -11.1 ^{+ 0.4 }_{ -0.4 } $& $ -19.5 ^{+ 0.3 }_{ -0.3 } $& A & 1 \\
Gaia DR3 3046020085121862656 & 105.889456 & -11.57415 & $ -4.507 \pm 0.036 $& $ 1.473 \pm 0.033 $& $ 0.846 \pm 0.035 $& $ 26.2 \pm 1.4 $& $ 1157.3 ^{+ 0.7 }_{ -0.7 } $& $ -33.2 ^{+ 0.5 }_{ -0.5 } $& $ -6.8 ^{+ 0.5 }_{ -0.5 } $& $ -19.6 ^{+ 0.4 }_{ -0.4 } $& A & 1 \\
Gaia DR3 3046018882531291264 & 105.912736 & -11.62173 & $ -4.630 \pm 0.081 $& $ 1.435 \pm 0.080 $& $ 0.830 \pm 0.080 $& $ 31.3 \pm 1.1 $& $ 1178.3 ^{+ 1.7 }_{ -1.7 } $& $ -34.7 ^{+ 1.0 }_{ -1.1 } $& $ -7.8 ^{+ 1.1 }_{ -1.1 } $& $ -20.5 ^{+ 0.9 }_{ -0.9 } $& A & 1 \\
Gaia DR3 3046019294847876864 & 105.925358 & -11.58721 & $ -4.425 \pm 0.035 $& $ 1.545 \pm 0.032 $& $ 0.779 \pm 0.033 $& $ 34.1 \pm 1.6 $& $ 1254.2 ^{+ 0.7 }_{ -0.7 } $& $ -36.4 ^{+ 0.5 }_{ -0.5 } $& $ -7.6 ^{+ 0.5 }_{ -0.5 } $& $ -20.6 ^{+ 0.5 }_{ -0.4 } $& A & 1 \\
Gaia DR3 3046027021488136960 & 105.939816 & -11.47577 & $ -4.399 \pm 0.082 $& $ 1.615 \pm 0.082 $& $ 0.756 \pm 0.081 $& $ 30.0 \pm 0.5 $& $ 1288.3 ^{+ 1.8 }_{ -1.8 } $& $ -35.7 ^{+ 0.9 }_{ -1.0 } $& $ -5.8 ^{+ 1.0 }_{ -0.9 } $& $ -20.6 ^{+ 1.0 }_{ -1.0 } $& A & 1 \\
Gaia DR3 3046026853990282496 & 105.940954 & -11.49571 & $ -3.941 \pm 0.044 $& $ 1.696 \pm 0.041 $& $ 0.886 \pm 0.043 $& $ 31.2 \pm 0.8 $& $ 1109.0 ^{+ 0.8 }_{ -0.9 } $& $ -33.5 ^{+ 0.6 }_{ -0.6 } $& $ -8.7 ^{+ 0.6 }_{ -0.6 } $& $ -15.7 ^{+ 0.5 }_{ -0.6 } $& A & 1 \\
Gaia DR3 3046019913323142400 & 105.958073 & -11.53746 & $ -4.516 \pm 0.032 $& $ 1.664 \pm 0.031 $& $ 0.884 \pm 0.032 $& $ 35.6 \pm 1.6 $& $ 1111.1 ^{+ 0.8 }_{ -0.7 } $& $ -34.9 ^{+ 0.4 }_{ -0.5 } $& $ -8.4 ^{+ 0.5 }_{ -0.5 } $& $ -18.5 ^{+ 0.4 }_{ -0.4 } $& A & 1 \\
Gaia DR3 3045831274058261120 & 105.963795 & -11.62890 & $ -4.480 \pm 0.104 $& $ 1.601 \pm 0.109 $& $ 0.779 \pm 0.110 $& $ 34.0 \pm 0.7 $& $ 1252.6 ^{+ 3.3 }_{ -3.3 } $& $ -37.2 ^{+ 1.2 }_{ -1.1 } $& $ -8.2 ^{+ 1.2 }_{ -1.2 } $& $ -20.7 ^{+ 1.3 }_{ -1.1 } $& A & 1 \\
Gaia DR3 3046019604085794560 & 105.967688 & -11.57208 & $ -4.552 \pm 0.066 $& $ 1.636 \pm 0.066 $& $ 0.829 \pm 0.069 $& $ 26.1 \pm 0.7 $& $ 1177.6 ^{+ 1.7 }_{ -1.6 } $& $ -33.0 ^{+ 0.8 }_{ -0.9 } $& $ -5.0 ^{+ 0.9 }_{ -0.8 } $& $ -19.5 ^{+ 0.8 }_{ -0.8 } $& A & 1 \\
Gaia DR3 3046026750911057408 & 105.970952 & -11.49312 & $ -4.620 \pm 0.020 $& $ 1.653 \pm 0.019 $& $ 0.855 \pm 0.019 $& $ 28.3 \pm 0.6 $& $ 1146.0 ^{+ 0.4 }_{ -0.4 } $& $ -33.6 ^{+ 0.3 }_{ -0.3 } $& $ -5.9 ^{+ 0.3 }_{ -0.3 } $& $ -19.4 ^{+ 0.2 }_{ -0.2 } $& A & 1 \\
Gaia DR3 3046019531067048320 & 105.972419 & -11.58067 & $ -4.610 \pm 0.032 $& $ 1.670 \pm 0.031 $& $ 0.861 \pm 0.030 $& $ 27.6 \pm 0.6 $& $ 1138.9 ^{+ 0.7 }_{ -0.8 } $& $ -33.3 ^{+ 0.5 }_{ -0.5 } $& $ -5.7 ^{+ 0.4 }_{ -0.5 } $& $ -19.2 ^{+ 0.3 }_{ -0.4 } $& A & 1 \\
Gaia DR3 3045831759397198080 & 105.977989 & -11.60158 & $ -4.738 \pm 0.067 $& $ 1.462 \pm 0.066 $& $ 0.908 \pm 0.070 $& $ 36.0 \pm 0.8 $& $ 1086.3 ^{+ 1.7 }_{ -1.7 } $& $ -35.7 ^{+ 0.8 }_{ -0.8 } $& $ -10.3 ^{+ 0.9 }_{ -0.8 } $& $ -19.6 ^{+ 0.8 }_{ -0.7 } $& A & 1 \\
Gaia DR3 3045830208906315392 & 105.985779 & -11.65612 & $ -4.444 \pm 0.085 $& $ 1.578 \pm 0.094 $& $ 0.799 \pm 0.084 $& $ 39.9 \pm 0.7 $& $ 1227.0 ^{+ 2.0 }_{ -2.0 } $& $ -39.3 ^{+ 1.1 }_{ -1.2 } $& $ -11.1 ^{+ 1.2 }_{ -1.1 } $& $ -20.4 ^{+ 1.0 }_{ -1.0 } $& A & 1 \\
Gaia DR3 3045829968388105472 & 106.010076 & -11.67094 & $ -4.377 \pm 0.090 $& $ 1.608 \pm 0.094 $& $ 0.777 \pm 0.096 $& $ 31.6 \pm 1.4 $& $ 1256.6 ^{+ 2.2 }_{ -2.1 } $& $ -35.9 ^{+ 1.3 }_{ -1.3 } $& $ -7.2 ^{+ 1.3 }_{ -1.2 } $& $ -20.1 ^{+ 1.0 }_{ -1.1 } $& A & 1 \\
Gaia DR3 3046032179749668992 & 106.019563 & -11.39436 & $ -4.641 \pm 0.081 $& $ 1.668 \pm 0.087 $& $ 0.803 \pm 0.076 $& $ 41.5 \pm 1.1 $& $ 1220.6 ^{+ 1.6 }_{ -1.7 } $& $ -39.1 ^{+ 1.2 }_{ -1.2 } $& $ -9.5 ^{+ 1.2 }_{ -1.2 } $& $ -20.9 ^{+ 0.9 }_{ -0.9 } $& A & 1 \\
Gaia DR3 3045831621959430912 & 106.025279 & -11.59791 & $ -4.407 \pm 0.050 $& $ 1.548 \pm 0.056 $& $ 0.933 \pm 0.053 $& $ 30.9 \pm 0.6 $& $ 1055.7 ^{+ 1.2 }_{ -1.2 } $& $ -33.0 ^{+ 0.7 }_{ -0.7 } $& $ -8.9 ^{+ 0.7 }_{ -0.7 } $& $ -17.4 ^{+ 0.5 }_{ -0.5 } $& A & 1 \\
Gaia DR3 3046033829017076352 & 106.027389 & -11.30965 & $ -4.881 \pm 0.036 $& $ 1.628 \pm 0.040 $& $ 0.856 \pm 0.034 $& $ 39.2 \pm 0.9 $& $ 1146.4 ^{+ 0.8 }_{ -0.8 } $& $ -38.1 ^{+ 0.6 }_{ -0.6 } $& $ -9.5 ^{+ 0.7 }_{ -0.6 } $& $ -20.9 ^{+ 0.4 }_{ -0.4 } $& A & 1 \\

\hline\hline
\end{tabular}
}
\tablefoot{We provide for each star Gaia DR3 identifier, position, proper motion, parallax, radial velocity measured in this study, Bayesian distance derived in this study, $UVW$ velocity components, subgroup of the CMa region, and flag indicating whether star was retained in our analysis after outlier rejection in velocity space (0=rejected, 1= accepted). This table will be available in its entirety in machine-readable form at the CDS.}
\end{table}
\end{landscape}

\end{appendix}
\end{document}